\documentclass[journal]{IEEEtran}
\usepackage{amsmath}
\usepackage{amsfonts,amssymb}
\usepackage{graphicx}
\allowdisplaybreaks[4]
\usepackage{xcolor}
\definecolor{deepred}{RGB}{205,38,38}
\usepackage{algorithm}
\usepackage{algorithmic}
\usepackage{amsthm}

\usepackage{lineno,hyperref}
\modulolinenumbers[5]
\usepackage{bm}

\usepackage{multirow}
\usepackage{array}
\usepackage{booktabs}
\usepackage{lipsum}
\usepackage{xcolor}
\usepackage{etoolbox}

\ifCLASSINFOpdf

\else

\fi

\ifCLASSOPTIONcompsoc
  \usepackage[caption=false,font=normalsize,labelfont=sf,textfont=sf]{subfig}
\else
  \usepackage[caption=false,font=footnotesize]{subfig}
\fi

\begin{document}
\bibliographystyle{ieeetr} 

\title{Online Voltage Control for Unbalanced Distribution Networks Using  Projected Newton Method}
\author{\IEEEauthorblockN{Rui Cheng, \textit{Graduate Student Member}, \textit{IEEE}, Zhaoyu Wang, \textit{Senior Member}, \textit{IEEE}, Yifei Guo, \textit{Member},  \textit{IEEE}, Qianzhi Zhang, \textit{Graduate Student Member},  \textit{IEEE}
}
\thanks{This work was supported in part by the U.S. Department of Energy Wind Energy Technologies Office under Grant DE-EE0008956, and in part by the National Science Foundation under ECCS 1929975 (Corresponding author: Zhaoyu Wang).}
\thanks{Rui Cheng, Zhaoyu Wang, Yifei Guo and Qianzhi Zhang are with the Department of Electrical and Computer Engineering, Iowa State University, Ames, IA 50011 USA (e-mail: ruicheng@iastate.edu; wzy@iastate.edu; yfguo.sdu@gmail.com; qianzhi@iastate.edu).}
}
\maketitle

\begin{abstract} 
This paper proposes an online voltage control strategy of distributed energy resources (DERs), based on the projected Newton method (PNM), for unbalanced distribution networks. The optimal Volt/VAr control (VVC) problem is formulated as an optimization program, with the goal of maintaining the voltage profile across the network by coordinating the VAr outputs of DERs. To overcome the slow convergence rate of conventional gradient-based methods, a PNM-based solution algorithm is developed to solve this VVC problem. It utilizes a non-diagonal
symmetric positive definite matrix, developed from the Hessian matrix of the objective, to scale the gradient, and thus a fast convergence performance can be expected in this Newton-like algorithm. Moreover, taking advantage of the instantaneous feedback of voltage measurements, the online implementation of the PNM-based VVC is further designed to deal with fast system variations. In this online PNM-based VVC scheme, each bus agent communicates the instantaneous voltage measurements to the central agent, and the central agent communicates the VAr output commands of DERs back to each bus agent. The fast convergence performance of PNM results in a stronger capability to track the system variations in real time. Finally, numerical case studies are performed to validate the effectiveness, superiority, and scalability of the proposed method.
\end{abstract}


\section{Introduction}
\IEEEPARstart{T}{he} increasing proliferation of distributed energy resources (DERs), such as photovoltaic (PV) generators, has posed new challenges to Volt/VAr control (VVC) problems in distribution networks due to the intermittent nature of renewable energy resources. However, meanwhile, it also provides promising opportunities to utilize DERs to resolve the VVC problems due to the great advance in the inverter-based technologies of DERs \cite{KT}.

The VVC problems are often cast as optimization programs with various control goals. {Some studies have made efforts to explore different VVC methods to coordinate DERs in distribution networks.} In \cite{MF}, the VVC problem is formulated as an optimal power flow (OPF) problem through a Second-Order Cone Program to minimize line losses and energy consumption in single-phase radial distribution networks. A sequential convex programming is applied to minimize total injected reactive power while satisfying the operational constraints of unbalanced distribution networks \cite{SD}. {However, the works \cite{MF,SD} are solved in the centralized manner, entailing large amounts of computational burdens and time.}

{Considering the high proliferation of DERs in distribution networks, various distributed VVC methods have attracted growing attentions, which can  be roughly classified into two categories: hierarchical and decentralized VVC strategies. Hierarchical VVC strategies establish on the communication between the central agent and local agents. In \cite{XZ}-\cite{YG2}, the hierarchical VVC strategies, based on the Alternating Direction Method of Multipliers (ADMM), are applied to coordinate the electric vehicle charging and wind turbines, respectively. In \cite{ED2}, an ADMM-based hierarchical management strategy is developed between the utility and customers to optimally dispatch  real
and reactive power set-points for residential PV inverters. In contrast, decentralized VVC strategies mainly rely on the communication between neighboring buses. The study by Zhang et al. \cite{BZ} proposes a dual decomposition-based decentralized VVC method to minimize power losses in single-phase networks, relying on the utilization of reactive-power-capable DERs. The authors in \cite{BAR} propose a voltage-constrained ADMM-based algorithm that optimally coordinates the VAr outputs of DERs in unbalanced distribution systems to regulate bus voltages. The ADMM-based strategies are utilized in \cite{PS}-\cite{WZ} to solve VVC problems as well.
} 
Those VVC problems in \cite{MF}-\cite{WZ} are solved in an \textit{offline} manner. That is, the solution in the offline voltage control cannot be applied to distribution networks till the iteration convergences. Thus, they might not be able to capture the fast fluctuations and disturbances in distribution networks caused by the high variability of DERs, potentially leading to undesirable and unacceptable results.

In recent years, distribution networks are undergoing a fundamental architecture transformation to become more intelligent, controllable, and open due to  the increasing deployment of communication, computing, information devices across distribution networks, such as advanced metering infrastructure. It offers an unprecedented promising opportunity to transform the voltage control scheme from the open-loop \textit{offline} manner to the closed-loop \textit{online} manner.

Different than the offline algorithms, the online algorithms naturally track changing network conditions as these changes manifest themselves in the network state that is used to compute the control solution \cite{OV2}. {The online VVC strategy has already triggered a wide-spread interest. One category is the traditional ``droop'' control \cite{MF2}-\cite{1547-2018}, which actively adjusts its VAr output as a function of local bus voltage. However, the work \cite{NL} points out that the droop control cannot maintain a feasible voltage profile under certain circumstances. Consequently, other more sophisticated VVC strategies have been proposed.} The authors in \cite{NL}-\cite{JL} propose various online VVC schemes based on a dual ascent method. In \cite{GQ}, a primal-dual gradient algorithm is leveraged to regulate voltage in real time through the iterative update of primal and dual variables. A gradient projection (GP) method is applied to design online voltage control with the goal of minimizing the voltage deviation in \cite{YG}. {Liu et al. \cite{HJL} propose an ADMM-based VVC strategy in an online fashion to solve voltage regulation problems, subject to time-varying operating conditions. Those underlying methodologies in online VVC strategies \cite{NL}-\cite{HJL} can be regarded as first-order gradient-based algorithms. A known drawback of gradient-based algorithms is that they may suffer from a slow convergence rate and are highly affected by the condition number of the problem \cite{BertsekasNP}. However, the convergence rate of algorithms is of great importance for the online implementation since it significantly affects the tracking capabilities. Researchers and practitioners have developed some newly online VVC strategies to improve the tracking capabilities. An accelerated dual descent algorithm is proposed to solve the voltage regulation problem in single-phase networks \cite{ZT} by employing the dual decomposition and accelerated gradient projected techniques. But this work is established on the assumption that the single-phase network lines have
the same ratio of resistance to reactance. 
The works \cite{HZ,WJ} propose feedback-based VVC strategies by means of a diagonally scaled gradient projection (DSGP) method, adopting a diagonal scaling of gradient to improve the convergence performance; however, the DSGP method theoretically only has a linear convergence rate \cite{ProjectedGradient}. {Moreover, although the simulations of \cite{HZ} are tested on unbalanced distribution networks, the theoretical analysis of \cite{HZ} only builds on single-phase distribution networks.} The Newton's method has always been a popular option to improve the convergence performance by directly scaling the gradient with the inverse Hessian matrix of the objective. However, using the inverse Hessian matrix directly could be problematic for constrained optimization problems. As indicated in \cite[Sec. 2.4]{BertsekasNP}, the Newton's update may fail to attain the optimal solution of a constrained program. A real-time OPF algorithm, based on quasi-Newton methods, is developed in \cite{YT} for single-phase networks. Nevertheless, it requires solving sub-optimization programs in the online implementation, thus its effectiveness is dependent upon the computational burdens of solving sub-optimization programs.}

{In this context, an online VVC scheme, based on the projected Newton method (PNM) \cite{ProjectedNewton}, is proposed to adjust the VAr outputs of DERs to maintain the voltage profile across unbalanced distribution networks. In this online PNM-based scheme, each bus agent sends the local voltage measurement as feedback to the central agent; then, the central agent communicates the VAr commands back to each agent to optimally coordinate the DERs across the unbalanced distribution network. The main contributions of this paper are summarized as follows:} 
\begin{itemize}
    \item The second-order PNM is applied to solve the constrained VVC problem with the goal of maintaining the voltage profile. Instead of directly using the inverse Hessian matrix of the objective, PNM adopts a non-diagonal positive definite symmetric matrix, developed from the Hessian matrix, to scale the gradient so that the resulting Newton-like algorithm can guarantee a fast convergence rate, {
   contributing to a great capability to track the time-varying changes in real time.} 
    \item {Taking advantage of the instantaneous local voltage measurement from each bus agent, an online feedback-based PNM-based hierarchical VVC strategy is further developed. Each local bus agent sends its local voltage measurement to the central agent, and the central agent communicates back to each bus agent the VAr output command. Under this online VVC strategy, the central agent and bus agents are not required to solve any (sub) optimization program. This attribute can effectively reduce the computational burdens, facilitating the online implementation to cope with fast system variations.
    }
    \item {The theoretical analysis and simulations are both built on unbalanced distribution networks.} We mathematically analyze this PNM-based VVC scheme, based on the linearized power flow model \cite{LG} of unbalanced distribution networks. And we test the performance of this PNM-based VVC scheme on the multi-phase nonlinear power flow in numerical case studies. Numerical comparisons with other VVC strategies are also provided to demonstrate the superiority of our proposed method.
\end{itemize}
Remaining sections are organized as follows.  The system modeling and problem statement are presented in Section \ref{sec:SystemModeling}. The details of leveraging PNM to solve the VVC problem are discussed in Section \ref{sec:VoltageControl}. Section \ref{sec:OnlineImplementation} outlines the online implementation strategy of the proposed PNM-based VVC. Numerical case studies and results are provided in Section \ref{sec:CaseStudy}. Concluding comments are given in Section
\ref{sec:Conclusion}.

\section{System Modeling and Problem Statement}\label{sec:SystemModeling}
\subsection{Unbalanced Radial Distribution Networks}
Consider an unbalanced radial distribution network with $N$+1 buses.  Let $\{0\}\bigcup\mathcal{N}$ denote the index set for these buses, where ${\mathcal{N}} = \{1,2,...,N\}$, bus 0 is the feeder head bus. For each bus $i\in\mathcal{N}$, let $bp(i)$ denote the bus that immediately
precedes bus $i$ along the radial network headed by bus 0. Let $\mathcal{L}=\{\ell_j=(i,j)|i=bp(j), j\in\mathcal{N}\}$ denote this edge set of line segments. Also, let $\mathcal{N}_i$ denote the set of all buses located strictly after bus $i$ along the radial network. 

 
 \textcolor{blue}{
\begin{table}[t]
    \centering
    \caption{Nomenclature: Operator}
    \label{tab:Nomenclature}
    \begin{tabular}{ll}
    \hline
        $\text{diag}(\bm{u})$ & A diagonal matrix with the entries of $\bm{u}$ in its diagonal\\
        $\bm{u}^{H}$ & The conjugate transpose of $\bm{u}$\\
        $||\bm{\mu}||_{\bm{C}}^2$ & It denotes $\bm{\mu}^T\bm{C}\bm{\mu}$ \\
        $\text{[}\bm{\mu}\text{]}_i$ &  For vector $\bm{\mu}$, let $[\bm{\mu}]_i$ denote the $i$-th element in $\bm{\mu}$\\
        $\text{[}\bm{U}\text{]}_{ij}$ & For matrix $\bm{U}$, let $[\bm{U}]_{ij}$ denote the $i$-th row and $j$-th column \\&element in $\bm{U}$\\
        \text{[} $\cdot$ \text{]}$_{\bm{\underline{q}}^g}^{\bm{\overline{q}}^g}$& Projection operator onto the box constraint $[\bm{\underline{q}}^g,\bm{\overline{q}}^g]$\\
        $\odot$& Element-wise multiplication\\ 
        $\oslash$&Element-wise division\\
    \hline
    \end{tabular}
\end{table}
}

\begin{figure}[ht]
     \centering
     \includegraphics[width=3in]{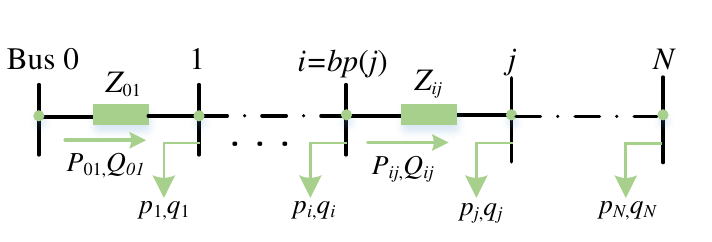}
     \caption{A radial distribution network.}
     \label{fig:RadialNetwork}
 \end{figure}

Let $\Phi_{i}$ and $n_i$ denote the phase set and number of bus $i\in\{0\}\bigcup\mathcal{N}$, $\Phi_{ij}$ and $n_{\ell_j}$ denote the phase set and number of line segment $\ell_j=(i,j)\in\mathcal{L}$. Note that for $\forall{\ell_j}\in\mathcal{L}$, $n_{\ell_j}=n_j$. For each bus $i\in\{0\}\bigcup\mathcal{N}$, $V_i^{\phi}(t)$ and $v_i^{\phi}(t)$ denote its phase $\phi$ complex voltage and squared voltage magnitude, respectively, and define the following column vectors: $\bm{V}_i(t)=[V_i^{\phi}(t)]_{\phi\in\Phi_i}\in\mathbb{C}^{n_i}$, $\bm{V}_i^{\Phi_{ij}}(t)=[V_i^{\phi}(t)]_{\phi\in\Phi_{ij}}\in\mathbb{C}^{n_{\ell_{j}}}$,
$\bm{v}_i(t)=[v_i^{\phi}(t)]_{\phi\in\Phi_i}\in\mathbb{R}^{n_i}$, $\bm{v}_i^{\Phi_{ij}}(t)=[v_i^{\phi}(t)]_{\phi\in\Phi_{ij}}\in\mathbb{R}^{n_{\ell_{j}}}$. And let column vectors $\bm{p}_i(t)=[p_i^{\phi}(t)]_{\phi\in\Phi_i}$ and $\bm{q}_i(t)=[q_i^{\phi}(t)]_{\phi\in\Phi_i}\in\mathbb{R}^{n_i}$ denote the real and reactive power consumption at bus $i$. For each line segment $(i,j)\in\mathcal{L}$, let $\bm{P}_{ij}(t)=[P_{ij}^{\phi}(t)]_{\phi\in\Phi_{ij}}$ and $\bm{Q}_{ij}(t)=[Q_{ij}^{\phi}(t)]_{\phi\in\Phi_{ij}}\in\mathbb{R}^{n_{\ell_j}}$ denote the real and reactive power flow from bus $i$ to $j$, and $\bm{Z}_{ij}\in\mathbb{C}^{n_{\ell_j}\times{n}_{\ell_j}}$ denote the impedance matrix for line segment $(i,j)$. {The nonlinear distribution power flow can be represented as follows, for $\forall{(i,j)}\in\mathcal{L}$,}
\begin{equation}\label{eq:nonlinearpf}
    \bm{V}_j(t)=\bm{V}_i^{\Phi_{ij}}(t)-\bm{Z}_{ij}[(\bm{P}_{ij}(t)-\bm{Q}_{ij}(t))\oslash\bm{V}_i^{\Phi_{ij}}(t)]
\end{equation}
{The nonlinear distribution power flow (\ref{eq:nonlinearpf}) poses challenges for optimization problems due to its nonconvex nature. Some convex relaxation approaches, e.g., semidefinite programming relaxation (\cite{LG,ED}), are leveraged to convexify the nonlinear distribution power flow. However, those convex relaxation approaches may be computationally expensive, and the exactness of convex relaxation may not be guaranteed. Instead, the linearized distribution power flow (\cite{BAR,LG}) exhibits lower complexity and higher computational efficiency by assuming (i) voltages among phases are nearly balanced; (ii) line losses are small. Compared to the nonlinear distribution power flow and  convex relaxation approaches, the linearized power flow model is easier to facilitate the algorithm design in optimization problems due to its linear and convex characteristics. The linearized distribution power flow generalizes the LinDistFlow equations \cite{MBF} from single-phase networks to multiphase networks.\footnote{{We refer readers to \cite{BAR,LG} for more details regarding the linearized distribution power flow. }}}  
The linearized distribution power flow can be presented as follows, for $\forall{(i,j)}\in\mathcal{L}$,
\begin{subequations}\label{eq:PowerFlow}
\begin{align}
    \bm{v}_j(t)&=\bm{v}_i^{{\Phi}_{ij}}(t)-2(\bm{\tilde{R}}_{ij}\bm{P}_{ij}(t)+\bm{\tilde{X}}_{ij}\bm{Q}_{ij}(t))\\
    \bm{P}_{ij}(t)&=\sum_{k\in{N}_{j}}\bm{P}_{jk}(t)+\bm{p}_j(t)\\
    \bm{Q}_{ij}(t)&=\sum_{k\in{N}_{j}}\bm{Q}_{jk}(t)+\bm{q}_j(t)
\end{align}
\end{subequations}
with
\begin{subequations}
\begin{align}
    \bm{\tilde{Z}}_{ij}&=\bm{\tilde{R}}_{ij}+j\bm{\tilde{X}}_{ij}=\Big[\big(\bm{a}\bm{a}^H\big)^{{\Phi}_{ij}}\odot\bm{Z}_{ij}^*\Big]^*\\
    \bm{a}&=[1,e^{-j\frac{2\pi}{3}},e^{j\frac{2\pi}{3}}]^T\\
    \bm{a}^H&=[1,e^{j\frac{2\pi}{3}},e^{-j\frac{2\pi}{3}}]
\end{align}
\end{subequations}
where $(\bm{a}\bm{a}^H)^{\Phi_{ij}}\in\mathbb{Z}^{n_{\ell_j}\times{n}_{\ell_j}}$ denotes the sub-matrix of $\bm{a}\bm{a}^H$, consisting of entries associated with $\Phi_{ij}$.
\subsection{Compact Form of Power Flow Model}
For an unbalanced radial distribution network, let  $m=\sum_{j=1}^Nn_{\ell_j}=\sum_{i=1}^N{n}_i$. Let $\bm{\bar{A}}=[\bm{A}_0, \bm{A}^T]^T\in\mathbb{R}^{(n_{0}+m)\times{m}}$ be the incidence matrix of unbalanced radial distribution network, where $\bm{A}_{0}^T\in\mathbb{R}^{n_0\times{m}}$ represents the connection  structure between bus 0 and each of the line segments in $\mathcal{L}$, $\bm{A}\in\mathbb{R}^{m\times{m}}$ represents the connection  structure  between the remaining buses and each of the line segments in $\mathcal{L}$ \footnote{Note that $\bm{\bar{A}}$ is invertible, see \cite[Appendix C]{RC} for details.}
. More precisely, the incidence matrix $\bm{\bar{A}}$ with an entry 1 for each “from”  phase node and -1 for each “to” phase node  corresponding to each phase circuit of line segments takes the following form:
\begin{equation}
\begin{split}
 \bm{\bar{A}}=
    \begin{bmatrix}
    \bm{J}(0,\ell_{1})&\bm{J}(0,\ell_{2})&...&\bm{J}(0,\ell_{N})\\
    \bm{J}(1,\ell_{1})&\bm{J}(1,\ell_{2})&...&\bm{J}(1,\ell_{N})\\
    \vdots&\vdots&\ddots&\vdots\\
     \bm{J}(N,\ell_{1})&\bm{J}(N,\ell_{2})&...&\bm{J}(N,\ell_{N})
    \end{bmatrix}   
\end{split}
\end{equation}
where $\bm{J}(i,\ell_{j})\in\mathbb{R}^{n_i\times{n}_{\ell_j}}$ indicates the connection structure between bus $i$ and line segment $\ell_j$; {see \cite[Appendix B]{RC} for one numerical example illustrating the construction of $\bm{\bar{A}}$ for an unbalanced radial network.}

Let the squared voltage magnitudes, real and reactive power consumption, and real and reactive power flows over line segments be denoted for the network by the following column vectors%
      \footnote{The squared voltage magnitudes and the real and reactive loads at buses $i$ are sorted in accordance 
                with the ordering of these buses from small to large $i$.  The real and reactive power flows over line segments $\ell_j$
                are sorted in accordance with the ordering of these line segments from small to large $j$.} :
$\bm{v}=[\bm{v}_i]_{i\in\mathcal{N}}$,  $\bm{p}=[\bm{p}_i]_{i\in\mathcal{N}}$,  $\bm{q}=[\bm{q}_i]_{i\in\mathcal{N}}$,  $\bm{P}=[\bm{P}_{bp(i)i}]_{bp(i)i\in\mathcal{L}}$,  $\bm{Q}=[\bm{Q}_{bp(i)i}]_{bp(i)i\in\mathcal{L}}$,
$\bm{v}_{0}=[v_{0}^a,v_{0}^{b},v_{0}^c]^T$.
The linearized distribution power flow model (\ref{eq:PowerFlow}) can be written in a compact form:
\begin{subequations}\label{eq:CompactPFForm}
\begin{align}
    -\bm{A}_0\bm{v}_0(t)-\bm{A}^T\bm{v}(t)&=-2\bm{D}_r\bm{P}(t)-2\bm{D}_x\bm{Q}(t)\\
    -\bm{A}\bm{P}(t)&=\bm{p}(t)\\
    -\bm{A}\bm{Q}(t)&=\bm{q}(t)
\end{align}
\end{subequations}
with
\begin{subequations}
\begin{align}
    \bm{D}_r&=\text{blkdiag}[\bm{\tilde{R}}_{bp(1)1},...,\bm{\tilde{R}}_{bp(N)N}]\\
    \bm{D}_x&=\text{blkdiag}[\bm{\tilde{X}}_{bp(1)1},...,\bm{\tilde{X}}_{bp(N)N}]
\end{align}
\end{subequations}
We separate $\bm{q}(t)$ into two parts $\bm{q}^g(t)$ and $\bm{q}^c(t)$, where $\bm{q}(t)=\bm{q}^c(t)-\bm{q}^g(t)$, $\bm{q}^g(t)$
denotes the vector of reactive power generated by DERs, e.g., PV inverters, and $\bm{q}^c(t)$ denotes the reactive power consumption vector generated by other resources except DERs. {Substituting $\bm{q}(t)=\bm{q}^c(t)-\bm{q}^g(t)$, (\ref{eq:CompactPFForm}b) and (\ref{eq:CompactPFForm}c) into (\ref{eq:CompactPFForm}a), then we have:}
\begin{equation}\label{eq:CompactLinearModel}
    \bm{v}(t)=\underbrace{2[\bm{A}^T]^{-1}\bm{D}_x\bm{A}^{-1}}_{\bm{M}}\bm{q}^{g}(t)+\bm{c}(t)
\end{equation}
with
\begin{equation}\label{eq:c}
\begin{split}
       \bm{c}(t)&=-\bm{M}\bm{q}^{c}(t)-[\bm{A}^T]^{-1}\bm{A}_0\bm{v}_0(t)\\&-2[\bm{A}^T]^{-1}\bm{D}_r\bm{A}^{-1}\bm{p}(t)
\end{split}
\end{equation}
where vector $\bm{c}(t)$ is a function of $\bm{v}_{0}(t),\bm{p}(t),\bm{q}^{c}(t)$, reflecting the impacts of slack bus, real power, and other reactive power except DERs.

\medskip
\noindent
\textbf{Remark 1:}  The linearized distribution power flow (\ref{eq:CompactLinearModel}) is adopted in this work since the structure of (\ref{eq:CompactLinearModel}) can facilitate the algorithm design and theoretical analysis. Note that our proposed voltage control scheme can also be applied to the multi-phase nonlinear power flow model. In our numerical case studies, we test the performance of our 
proposed voltage control scheme on the nonlinear model using the open-source simulator OpenDSS \cite{OpenDss}.


\subsection{Problem Statement}
The goal of voltage control problem is assumed to minimize the voltage deviations by optimally coordinating the VAr outputs of DERs \footnote{{For this VVC problem, we do not consider the hard voltage constraint, but instead treat the voltage constraint as a soft penalty in the objective function to facilitate the algorithm design, just like \cite{ZT,HZ}}.}, which can be formulated as follows:
\begin{subequations}\label{eq:ProblemStatement}
\begin{align}
    \min_{\bm{{q}}^g(t),\bm{v}(t)} &{\frac{1}{2}}||\bm{v}(t)-\bm{v}_r||_2^2\\
    \text{subject to:~} &\bm{\underline{q}}^g(t)\leq\bm{{q}}^g(t)\leq\bm{\overline{q}}^g(t)\\
    &\bm{v}(t)=\bm{M}\bm{q}^g(t)+\bm{c}(t)
\end{align}
\end{subequations}
where $\bm{v}_r\in\mathbb{R}^m$ is the reference of squared voltage magnitude,  (\ref{eq:ProblemStatement}b) is the VAr limits for DERs, {(\ref{eq:ProblemStatement}c) is the compact form of linearized distribution power flow constraints.}
\section{Voltage Control Using Projected Newton Method}
\label{sec:VoltageControl}

\subsection{Problem Reformulation}
Substituting (\ref{eq:ProblemStatement}c) into (\ref{eq:ProblemStatement}a), we have:
\begin{equation}\label{eq:HFunction}
    h(\bm{q}^g(t))=\frac{1}{2}||\bm{v}(t)-\bm{v}_r||_2^2=\frac{1}{2}||\bm{M}\bm{q}^{g}(t)+\bm{c}(t)-\bm{v}_r||_2^2
\end{equation}
Let $\mathcal{X}(t)=\{\bm{q}^g(t)~|\bm{\underline{q}}^g(t)\leq\bm{{q}}^g(t)\leq\bm{\overline{q}}^g(t)\}$.
Then, the problem (\ref{eq:ProblemStatement}) can be reformulated as follows:
\begin{equation}\label{eq:Reformulation}
    \min_{\bm{q}^g(t)\in\mathcal{X}(t)}h(\bm{q}^g(t))
\end{equation}

\medskip
\noindent
\textbf{Remark 2:} With respect to single-phase radial  distribution networks, the works \cite{YG} and \cite{HZ} adopt $||\cdot||_{\bm{M}^{-1}}^2$ to replace $||\cdot||_2^2$ in (\ref{eq:HFunction}) \footnote{$\bm{M}$ is a symmetric positive definite matrix in single-phase radial  distribution networks.}, which facilitates the local/decentralized voltage control design. There might be two limitations in this way. First, the surrogate VVC problem {with} $||\cdot||_{\bm{M}^{-1}}^2$ assigns different weights to the squared voltage magnitude deviations and their product, which is not physical interpretable. Second, it might be improper to replace $||\cdot||_2^2$ by $||\cdot||_{\bm{M}^{-1}}^2$ in unbalanced radial distribution networks.
Since $\bm{D}_x$ is not a symmetric matrix in unbalanced radial distribution networks, $\bm{M}$ is no longer a symmetric positive definite matrix in unbalanced radial distribution networks. 

\medskip
\noindent
\textbf{Proposition 1:} 
\textit{ Let $\bm{H}$ denote the symmetric Hessian matrix of (\ref{eq:HFunction}), where $\bm{H}=\nabla^2h(\bm{q}^g(t))=\bm{M}^T\bm{M}$. The Hessian matrix $\bm{H}$ is a symmetric positive definite matrix as $\bm{D}_x$ is invertible. 
}

\noindent
\medskip
\textbf{Proof of Proposition 1:}
$\bm{H}$ can be represented as follows:
\begin{equation}
\begin{split}
    \bm{H}=\bm{M}^T\bm{M}&=4\big(\bm{A}^{-T}\bm{D}_x^T\bm{A}^{-1}\big)\cdot\big(\bm{A}^{-T}\bm{D}_x\bm{A}^{-1}\big)\\
    &=4\bm{A}^{-T}\bm{D}_x^T\cdot\big(\bm{A}^{-1}\bm{A}^{-T}\big)\cdot\bm{D}_x\bm{A}^{-1}
\end{split}
\end{equation}
For any non-zero row vector ${\bm{u}}$, we have:
\begin{equation}
    \bm{u}\bm{A}^{-1}\bm{A}^{-T}\bm{u}^T=\big(\bm{u}\bm{A}^{-1}\big)\cdot\big(\bm{u}\bm{A}^{-1}\big)^T>0
\end{equation}
It means that $\bm{A}^{-1}\bm{A}^{-T}$ is positive definite. Then, we have:
\begin{equation}
\begin{split}
    &\bm{u}(\bm{M}^T\bm{M})\bm{u}^T=\\&4\big(\bm{u}\bm{A}^{-T}\bm{D}_x^T\big)\cdot\big(\bm{A}^{-1}\bm{A}^{-T}\big)\cdot\big(\bm{u}\bm{A}^{-T}\bm{D}_x^T\big)^T
\end{split}
\end{equation}
As $\bm{D}_x$ is invertible, $\bm{u}\bm{A}^{-T}\bm{D}_x^T$ is a non-zero row vector.
Plus $\bm{A}^{-1}\bm{A}^{-T}$ is positive definite, it follows that $\bm{u}(\bm{M}^T\bm{M})\bm{u}^T>0$ which implies that $\bm{M}^T\bm{M}$ is positive definite. 
Plus $\bm{H}^T=\bm{M}^T\bm{M}=\bm{H}$, it follows that $\bm{H}\in\mathbb{S}_{++}^m$ is a symmetric positive definite matrix. Q.E.D.

\subsection{Projected Newton Method}
In this subsection, we propose a voltage control scheme using the projected Newton method. To facilitate algorithm design
and theoretic analysis, we assume $\bm{c}(t)=\bm{c}$, $\bm{\underline{q}}^g(t)=\bm{\underline{q}}^g,\bm{\overline{q}}^g(t)=\bm{\overline{q}}^g$ are fixed. But this is not required when applying our method, the online implementation of our proposed method is discussed in Section \ref{sec:OnlineImplementation} to deal with the time-varying conditions. The problem (\ref{eq:Reformulation}) is a strictly convex problem with box constraints since the Hessian matrix $\bm{H}$ is positive definite (See Proposition 1). For this type of problem, it can be solved by the GP method, which iteratively updates the VAr outputs of DERs at step $t$ in the following manner:
\begin{equation}\label{eq:GP}
    \bm{q}^g(t+1)=[\bm{q}^g(t)-\nabla{h}(\bm{q}^g(t))]_{\bm{\underline{q}}^g}^{\bm{\overline{q}}^g}
\end{equation}
with
\begin{subequations}\label{eq:Gradient}
\begin{align}
    \nabla{h}(\bm{q}^g(t))&=\bm{M}\big[\bm{v}(\bm{q}^g(t))-\bm{v}_r\big]\\
    \bm{v}(\bm{q}^g(t))&=\bm{M}\bm{q}^{g}(t)+\bm{c}
\end{align}
\end{subequations}
where $[$ $\cdot$ $]_{\bm{\underline{q}}^g}^{\bm{\overline{q}}^g}$ denotes the projection operator onto the box constraint $[\bm{\underline{q}}^g,\bm{\overline{q}}^g]$. However, the condition number of $\bm{H}$ could be large, the gradient-based methods, e.g., GP, always suffer from the slow convergence rate. 

To resolve the slow convergence implementation dilemma, one possible method is to scale the gradient with some positive definitive matrices such that:
\begin{equation}\label{eq:ScaledGradient}
    \bm{q}^g(t+1)=[\bm{q}^g(t)-\alpha(t)\bm{u}(t)]_{\bm{\underline{q}}^g}^{\bm{\overline{q}}^g}
\end{equation}
with
\begin{equation}\label{eq:U(K)}
    \bm{u}(t)=\bm{D}(t)\nabla{h}(\bm{q}^g(t)) 
\end{equation}
where $\alpha(t)$ and $\bm{D}(t)$ are a positive step size and a positive definite matrix. The work \cite{HZ} makes use of a diagonal positive definite matrix $\bm{D}(t)$ to achieve a proper diagonal scaling of $\nabla{h}$. However, the convergence rate of DSGP method is still not ideal, which is typically characterized by the linear convergence rate \cite{ProjectedGradient}.
Any attempt to construct a superlinearly convergent algorithm should by necessity involve a nondiagonal scaling matrix. This motivates us to explore the use of more general, nondiagonal scaling of $\nabla{h}$. The Newton's method is always a popular approach by scaling the gradient with the inverse Hessian matrix of the objective. However, the scaling manner by directly using the inverse Hessian matrix could be problematic when it comes to constrained optimization problems with a projection operator. The iterated manner is not always in general a descent iteration. That is, we may have $h(\bm{q}^g(t+1))>h(\bm{q}^g(t))$ for all positive step size $\alpha(t)$ with an unfavorable choice of $\bm{D}(t)$ (see \cite[Sec.~2.4]{BertsekasNP} for details). In short, the scaling manner for constrained optimization problems is far more complicated than their unconstrained counterparts. {The appropriate selection of $\bm{D}(t)$ is of great importance for the scaling manner for constrained optimization problems.}{To this end, the following Proposition 2 identifies a class of matrices $\bm{D}(t)$ for which a descent iteration is obtained. Proposition 2 can be easily proved by using \cite[Proposition 1]{ProjectedNewton}.\\
\noindent
\textbf{Definition} ($\bm{D}$ is diagonal with respect to $\bm{I}$): If a symmetric $m\times{m}$ matrix $\bm{D}$ and a subset of indices $\bm{I}\subset\{1,2,...,m\}$ satisfy:
\begin{equation}
    [\bm{D}]_{ij}=0, \mbox{for~} i\in\bm{I}, j=1,2,...,m, i\neq{j}
\end{equation}
then we say that $\bm{D}$ is diagonal with respect to $\bm{I}$.}\\
{And define $\bm{\hat{I}}(t)$:
\begin{equation}
\begin{split}
    \bm{\hat{I}}(t)=&\Big\{i|[\bm{q}^{g}(t)]_i=[\bm{\underline{q}}^{g}]_i \text{~and~}\frac{\partial{h}(\bm{q}^g(t))}{\partial{[\bm{q}^{g}}]_i}>0;\\
    \text{or~}& [\bm{q}^{g}(t)]_i=[\bm{\overline{q}}^{g}]_{i} \text{~and~}\frac{\partial{h}(\bm{q}^g(t))}{\partial{[\bm{q}^{g}}]_i}<0
    \Big\}
\end{split}
\end{equation}}
\noindent
\textbf{Proposition 2}: Assume $\bm{D}(t)$ is a symmetric positive definite matrix which is diagonal with respect to $\bm{\hat{I}}(t)$, then we have:
\begin{itemize}
    \item \textbf{[P2.a]} If $\bm{q}^g(t)$ is the optimal solution $\bm{q}^{g\ast}$  of the problem (\ref{eq:Reformulation}), then:
    \begin{align*}
        \bm{q}^g(t+1)&=[\bm{q}^g(t)-\alpha(t)\bm{u}(t)]_{\bm{\underline{q}}^g}^{\bm{\overline{q}}^g}\\&=\bm{q}^g(t), \forall{\alpha(t)>0}
    \end{align*}
    \item \textbf{[P2.b]} If $\bm{q}^g(t)$ is not the optimal solution $\bm{q}^{g\ast}$  of the problem (\ref{eq:Reformulation}), then there exists a scalar $\bar{\alpha}(t)$ such that:
        \begin{align*}
        h(\bm{q}^g(t+1))<{h}(\bm{q}^g(t)), \forall{\alpha(t)}\in(0,\bar{\alpha}(t)]
    \end{align*}
    with
    \begin{align*}
        &\bm{q}^g(t+1)=[\bm{q}^g(t)-\alpha(t)\bm{u}(t)]_{\bm{\underline{q}}^g}^{\bm{\overline{q}}^g}\\
        &\bm{u}(t)=\bm{D}(t)\nabla{h}(\bm{q}^g(t)), \text{defined in }(\ref{eq:U(K)}).
    \end{align*}
\end{itemize}

{Proposition 2 shows that the iteration essentially terminates at a optimal solution and is capable of descent when
not at a global minimum. Unfortunately, $\bm{\hat{I}}(t)$ in Proposition 2 exhibits an undesirable discontinuity at the boundary of the constraint set, i.e., $\bm{\underline{q}}^g$ and $\bm{\overline{q}}^g$, which could have an adverse effect on its rate of convergence. (This phenomenon is quite common in feasible
direction algorithms and is referred to as zigzagging or jamming). Consequently, we aim to introduce  $\bm{\bm{I}(t)}$, certain enlargements of $\bm{\hat{I}}(t)$, to bypass those difficulties, where $\bm{I}(t)$ is defined as follows:
\begin{equation}\label{eq:I(k)}\small
    \begin{split}
\bm{I}(t)=&\Big\{i|{[\bm{\underline{q}}}^{g}]_i\leq[\bm{q}^{g}(t)]_i\leq[\bm{\underline{q}}^g]_i+[\bm{\varepsilon}(t)]_i \text{~and~}\frac{\partial{h}(\bm{q}^g(t))}{\partial{[\bm{q}^{g}}]_i}>0;\\
\text{or~}& [\bm{\overline{q}}^{g}]_i-[\bm{\varepsilon}(t)]_i\leq[\bm{q}^{g}(t)]_i\leq[\bm{\overline{q}}^{g}]_{i} \text{~and~}\frac{\partial{h}(\bm{q}^g(t))}{\partial{[\bm{q}^{g}}]_i}<0
\Big\}
    \end{split}
\end{equation}
where
\begin{equation}\label{eq:varepsilon}
    [\bm{\varepsilon}(t)]_i=\min\{\varepsilon,[\bm{w}(t)]_i\}
\end{equation}
$\varepsilon$ is a fixed positive scalar (typically small) and $[\bm{w}(t)]_i$ is the $i$-th element in $\bm{w}(t)$, $\bm{w}(t)$ is given by \begin{equation}\label{eq:w}
     \bm{w}(t)=\Big|\bm{q}^g(t)-[\bm{q}^g(t)-\bm{C}\nabla{h}(\bm{q}^g(t))]_{\bm{\underline{q}}^g}^{\bm{\overline{q}}^g}\Big|
\end{equation}
where $|$ $\cdot$ $|$ denotes the element-wise absolute value operation, $\bm{C}$ is a fixed diagonal positive definite matrix (e.g., the identity matrix).
}{This is an anti-zigzagging procedure commonly employed in feasible direction methods (see \cite{EP}), and is designed to counteract the possible discontinuity exhibited by the set $\bm{\hat{I}}(t)$. Ideally, $\bm{D}(t)$ is not only diagonal with respect to $\bm{\hat{I}}(t)$, but rather with respect to the larger set $\bm{{I}}(t)$. Meanwhile, to achieve a fast convergence performance, $\bm{D}(t)$ should be an adequate approximation of the inverse Hessian $\bm{H}^{-1}$, at least along a suitable subspace. At this point, Proposition 3 introduces  $\bm{D}(t)$ such that it is diagonal with respect $\bm{I}(t)$ and the portion of $\bm{D}(t)$ corresponding to the indices $i\notin\bm{I}(t)$ is the inverse of the Hessian $\bm{H}$ with respect to these indices.
}

\noindent
\textbf{Proposition 3:} Suppose $\bm{D}(t)$ satisfies:
\begin{equation}\label{eq:D}
    \bm{D}(t)=\bm{E}(t)^{-1}
\end{equation}
where $\bm{E}(t)$ is:
\begin{equation}\label{eq:E}\small
    \begin{split}
         [\bm{E}(t)]_{ij}&=
    \begin{cases}
     0 &\mbox{if~} i\neq{j}, \mbox{and either~} i\mbox{~or~} j\in\bm{I}(t)\\
     \big|[\bm{H}]_{ii}\big| & \mbox{if~} i=j \mbox{~and~} i\in\bm{I}(t)\\
     [\bm{H}]_{ij} & \mbox{otherwise}
    \end{cases}
    \end{split}
\end{equation}
Then, \textit{$\bm{D}(t)$ is a symmetric positive definite matrix, which is diagonal with respect to $\bm{I}(t)$.}

\noindent
\textbf{Proof of Proposition 3:} Without loss of generality, let $\gamma$ denote the number of elements in $\bm{I}(t)$. Define $\bm{\tilde{E}}(t)$ as follows:
\begin{equation}\label{eq:TildeE}
    \bm{\tilde{E}}(t)=
    \begin{bmatrix}
    \bm{B}(t)&&&\\
    &\big|[\bm{H}]_{I_{1}(t)I_{1}(t)}\big|&&\\
    &&\ddots&\\
    &&&\big|[\bm{H}]_{I_{\gamma(t)}{I_\gamma(t)}}\big|
    \end{bmatrix}
\end{equation}
where $\bm{B}(t)$ is a $(m-\gamma)\times(m-\gamma)$ principal sub-matrix of $\bm{H}$ formed by removing any $i$-th row and column for $i\in\bm{I}(t)$. And let $I_j(t)$ denote the $j$-th element in $\bm{I}(t)$, where $j=1,2,..,\gamma$.

Since $\bm{H}$ is positive definite, it follows that $\bm{B}(t)$ is positive definite since any principal sub-matrix of positive definite matrix is positive definite. Thus, $\bm{\tilde{E}}(t)$ is a positive definite matrix. For any vector $\bm{\nu}\in\mathbb{R}^{m}$, we have:
\begin{equation}
    \bm{\nu}^T\bm{E}(t)\bm{\nu}=\bm{\bar{\nu}}^T\tilde{\bm{E}}(t)\bm{\bar{\nu}}>{0}
\end{equation}
where $\bm{\bar{\nu}}$ is a vector formed by rearranging elements in $\bm{\nu}$ in the order consistent with $\bm{\tilde{E}}(t)$, the inequality holds since $\bm{\tilde{E}}(t)$ is positive definite. Thus, $\bm{E}(t)$ is positive definite, implying $\bm{E}(t)$ is invertible. In addition, $\bm{E}(t)$ is symmetric since $\bm{H}$ is symmetric. Then, we can conclude that $\bm{E}(t)$ is a symmetric positive definite matrix, it follows that  $\bm{D}(t)$ is a symmetric positive definite matrix. Besides, from (\ref{eq:D}) and (\ref{eq:E}), it follows that:
\begin{equation}
    [\bm{D}(t)]_{ij}=0, \mbox{if~} i\neq{j}, \mbox{and either~} i\mbox{~or~} j\in\bm{I}(t).
\end{equation}
Then,  $\bm{D}(t)$ is diagonal with respect to $\bm{I}(t)$. Q.E.D.


\noindent
\textbf{Proposition 4:} Suppose $\bm{D}(t)=\bm{E}(t)^{-1}$, we have
\begin{itemize}
    \item \textbf{[P4.a]} \textbf{[P2.a]} holds.
    \item \textbf{[P4.b]} \textbf{[P2.b]} holds.
\end{itemize}

\medskip
\noindent
\textbf{Proof of Proposition 4:} See Appendix A.

{As indicated in Propositions 3 and 4, they extend $\bm{\hat{I}}(t)$ in Proposition 2 to the larger set $\bm{{I}}(t)$. Moreover,  $\bm{D}(t)=\bm{E}(t)^{-1}$ shows a suitable approximation of the inverse Hessian to some extent, and a descent iteration can be obtained as $\bm{D}(t)=\bm{E}(t)^{-1}$ holds.
}

{
Finally, we propose an offline PNM-based VVC strategy, scaling the gradient with the nondiagonal symmetric positive definite matrix $\bm{D}(t)=\bm{E}(t)^{-1}$, instead of directly making use of the inverse Hessian $\bm{H}^{-1}$. The offline PNM-based VVC is presented in Algorithm 1. Regrading the choice of step size in \textbf{S4} of Algorithm 1, it could be viewed as a combination of the Armijo-like rule and the Armijo rule. The convergence proof of PNM is provided in \cite{ProjectedNewton}. 
}
\begin{algorithm}[t]
\renewcommand\baselinestretch{1}\selectfont
\small
\caption{Offline PNM-based VVC}
\begin{algorithmic}[0]\label{PNM}
\STATE \hspace{-3mm}{\bf For any time step} {$t$}: Alternately update variables by the following steps until convergence: 

\STATE\hspace{-1mm} {\bf S1:} Update $\nabla{h}(\bm{q}^g(t))$ by (\ref{eq:Gradient}).
\STATE\hspace{-1mm} {\bf S2:} Update $\bm{w}(t)$ by (\ref{eq:w}). 
\STATE\hspace{-1mm} {\bf S3:} Identify the set $\bm{I}(t)$ by  (\ref{eq:I(k)}) and (\ref{eq:varepsilon}).
\STATE\hspace{-1mm} {\bf S4:} Update $\bm{D}(t)$ by (\ref{eq:D}) and (\ref{eq:E}), which is diagonal with respect to $\bm{I}(t)$. 
\STATE\hspace{-1mm} {\bf S5:} Update $\bm{q}^g(t+1)$. Let $\bm{u}(t)=\bm{D}(t)\nabla{h}(\bm{q}^g(t))$. Starting at $\tau(t)=0$, repeat:
\begin{align*}
    \tau(t)&=\tau(t)+1, \alpha(t)=\beta^{\tau(t)}\\
    \bm{q}^g(t+1)&=[\bm{q}^g(t)-\alpha(t)\bm{u}(t)]_{\underline{\bm{q}}^g}^{{\overline{\bm{q}}^g}}
\end{align*}
until:
\begin{align*}
    &h(\bm{q}^g(t))-h(\bm{q}^g(t+1))\geq\delta\Big\{\beta^{\tau(t)}\sum_{i\not\in\bm{I}(t)}\frac{\partial{h}(\bm{q}^g(t))}{\partial[\bm{q}^g]_i}[\bm{u}(t)]_i\\
    &+\sum_{i\in\bm{I}(t)}\frac{\partial{h}(\bm{q}^g(t))}{\partial[\bm{q}^g]_{i}}\big[[\bm{q}^g(t)]_i-[\bm{q}^g(t+1)]_i\big]\Big\}
\end{align*}
where $\beta\in(0,1)$ and $\delta\in(0,0.5)$.
\STATE\hspace{-1mm} {\bf S6:} Update $t\leftarrow{t+1}$.
\end{algorithmic}
\end{algorithm}

\medskip
\noindent
\textbf{Remark 3:} Gradient-based methods, e.g., GP, are first-order methods, they are typically characterized by the linear convergence rate. As discussed before, the second-order Newton's method cannot be directly applied to solve constrained optimization problems. There are some extensions and variants \cite{JCD}-\cite{UMGP} of Newton's method to solve constrained optimization problems, which are capable of superlinear convergence. However, sub-optimization problems are required to be solved in these methods \cite{JCD}-\cite{UMGP} (which could be very time consuming for large-scale systems), thus rendering these methods impractical. In contrast, PNM can be utilized without solving any (sub)optimization problems, thereby reducing the associated computational burden. It is proved in \cite[Prop.~4]{ProjectedNewton} that the superlinear convergence rate could be achieved for PNM under mild assumptions, thus leading to a faster convergence performance compared with first-order methods.

\section{Online implementation}\label{sec:OnlineImplementation}
To better capture the time-varying characteristic of system,  the online implementation of our proposed voltage control scheme is proposed in this section.  {With respect to the online implementation, $\bm{c}(t)$, $\bm{\overline{q}}^g(t)$, and $\bm{\underline{q}}^g(t)$ do not need to be fixed, which could be time-varying.}
\subsection{Estimation of {c}(t)}
$\bm{c}(t)$ is a function of $\bm{v}_{0}(t)$, $\bm{p}(t)$, $\bm{q}^c(t)$, where $\bm{c}(t)$ can be calculated by (\ref{eq:c}) for the online implementation.
\subsection{Estimation of VAr Limits}
VAr limits $\bm{\overline{q}}^g(t)$ and $\bm{\underline{q}}^g(t)$ of DERs can be updated based on the inverter capacities and the instantaneous real power output of DERs. The online update of VAr limits is helpful to guarantee inverters of DERs operate within a secure range, especially for preventing inverters from overload.

\begin{algorithm}[t]
\renewcommand\baselinestretch{1}\selectfont
\small
\caption{Online PNM-based VVC}
\begin{algorithmic}[0]\label{OnlineVoltage}
\STATE \hspace{-3mm}{\bf For any time step $t$, repeat the following steps:} 
\STATE\hspace{-1mm} {\bf S1:} Each bus agent $i$ measures voltage magnitudes and updates $\bm{v}_{i}^m(t)$, and estimates the VAr limits $\bm{\overline{q}}_i^g(t+1)$ and   $\bm{\underline{q}}_i^g(t+1)$.
\STATE\hspace{-1mm} {\bf S2:} Each bus agent $i$ sends $\bm{v}_{i}^m(t)$, $\bm{\overline{q}}_i^g(t+1)$ and $\bm{\underline{q}}_i^g(t+1)$ to the central agent.
\STATE\hspace{-1mm} {\bf S3:} The central agent estimates $\bm{c}(t)$, $\nabla{h}(\bm{q}^g(t))$ by (\ref{eq:c}) and (\ref{eq:UpdateGradient}).
\STATE\hspace{-1mm} {\bf S4:} The central agent updates the VAr limits by setting $\bm{\overline{q}}^g=\bm{\overline{q}}^g(t+1)$, $\bm{\underline{q}}^g=\bm{\underline{q}}^g(t+1)$.
\STATE\hspace{-1mm} {\bf S5:} The central agent updates $\bm{w}(t)$ by (\ref{eq:w}).
\STATE\hspace{-1mm} {\bf S6:} The central agent identifies the set $\bm{I}(t)$ by  (\ref{eq:I(k)}) and (\ref{eq:varepsilon}).
\STATE\hspace{-1mm} {\bf S7:} The central agent updates $\bm{D}(t)$ by (\ref{eq:D}) and (\ref{eq:E}), which is diagonal with respect to $\bm{I}(t)$. 
\STATE {\bf S8:} The central agent updates $\bm{q}^g(t+1)$. Let $\bm{u}(t)=\bm{D}(t)\nabla{h}(\bm{q}^g(t))$.
Starting at $\tau(t)=0$, repeat:
\begin{align*}
    \tau(t)&=\tau(t)+1, \alpha(t)=\beta^{\tau(t)}\\
    \bm{q}^g(t+1)&=[\bm{q}^g(t)-\alpha(t)\bm{u}(t)]_{\underline{\bm{q}}^g}^{{\overline{\bm{q}}^g}}
\end{align*}
until:
\begin{align*}
    &h(\bm{q}^g(t))-\hat{h}(\bm{q}^g(t+1))\geq\delta\Big\{\beta^{\tau(t)}\sum_{i\not\in\bm{I}(t)}\frac{\partial{h}(\bm{q}^g(t))}{\partial[\bm{q}^g]_i}[\bm{u}(t)]_i\\
    &+\sum_{i\in\bm{I}(t)}\frac{\partial{h}(\bm{q}^g(t))}{\partial[\bm{q}^g]_{i}}\big[[\bm{q}^g(t)]_i-[\bm{q}^g(t+1)]_i\big]\Big\}
\end{align*}
where $\hat{h}(\bm{q}^g(t+1))=\frac{1}{2}||\bm{M}\bm{q}^{g}(t+1)+\bm{c}(t)-\bm{v}_r||_2^2$,
$\beta\in(0,1)$ and $\delta\in(0,0.5)$. Note that $\bm{c}(t)$ is used in $\hat{h}(\bm{q}^g(t+1))$ to keep consistent with $h(\bm{q}^g(t))$. Then the central agent sends $\bm{q}_i^g(t+1)$ to bus agent $i$.
\STATE\hspace{-1mm} {\bf S9:} Update $t\leftarrow{t+1}$.
\end{algorithmic}
\end{algorithm}

\subsection{Estimation of Gradient}
For the online implementation, the rule to update the gradient $\nabla{h}(\bm{q}^g(t))=\bm{M}\big[\bm{v}(\bm{q}^g(t))-\bm{v}_r\big]$ as shown in (\ref{eq:Gradient}) can be replaced by a feedback control law based on the voltage measurement, which is:
\begin{equation}\label{eq:UpdateGradient}
    \nabla{h}(\bm{q}^g(t))=\bm{M}\big[\bm{v}^m(t)-\bm{v}_r\big]
\end{equation}
where $\bm{v}^m(t)$ is the squared voltage measurement vector for time step $t$. {It indicates the gradient $\nabla{h}(\bm{q}^g(t))$ can be estimated by utilizing the local voltage measurement from each bus agent in the online implementation.}

\begin{figure}[t]
     \centering
     \includegraphics[width=3.4in]{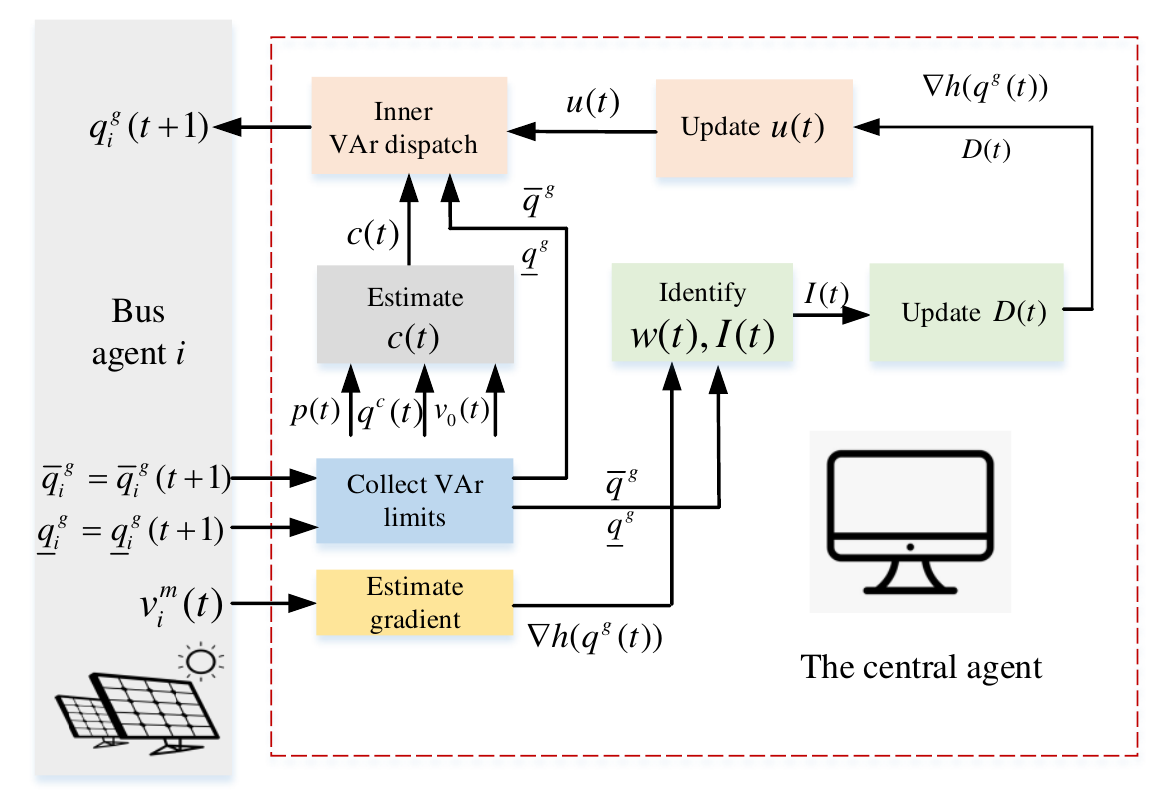}
     \caption{Online PNM-based VVC framework.}
     \label{fig:VArControl}
 \end{figure}
 \begin{figure}[t]
     \centering
     \includegraphics[width=3.2in]{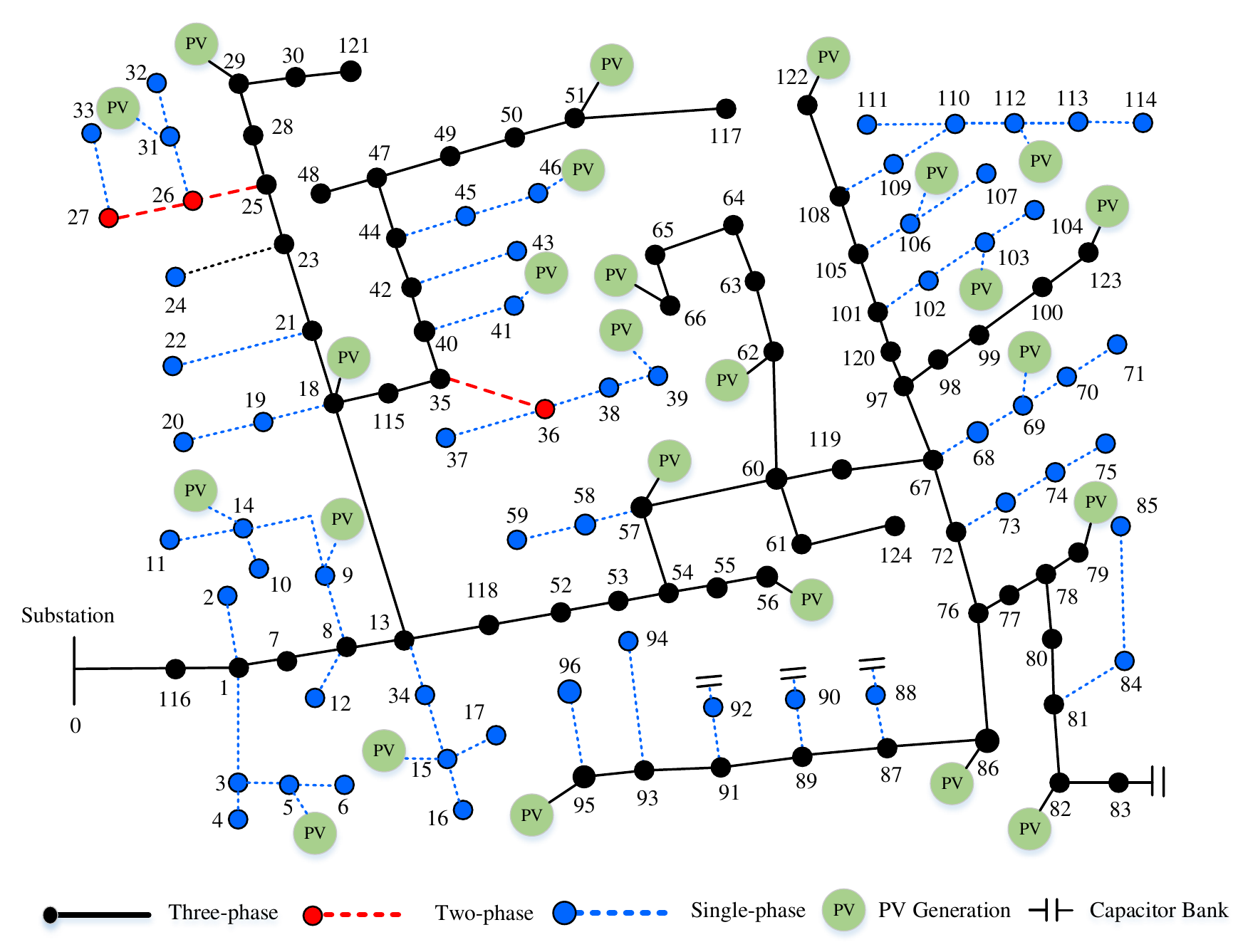}
     \caption{IEEE 123-bus test system.}
     \label{fig:IEEE123}
 \end{figure}
\subsection{Online PNM-based VVC}
The details and framework of the online PNM-based VVC are provided in Algorithm 2 and Fig.\ref{fig:VArControl}, respectively. As seen in Algorithm 2 and Fig.\ref{fig:VArControl}, each bus agent $i$ sends its voltage measurement $\bm{v}_i^m(t)$ and VAr limits to the central agent, and the central agent communicates the VAr output command $\bm{q}_i^g(t+1)$ back to each bus agent $i$. Note that there is no any (sub)optimization problem required to solve for both the central agent and local bus agent, the central agent only needs to implement the analytical calculation for each time step, such characteristic is suitable for online implementation. Moreover, the fast convergence performance of PNM contributes to a better capability to track the changes in distribution networks.

\medskip
\noindent
\textbf{Remark 4:} The key feature of online implementation is its own closed-loop nature, exploiting the most up-to-date voltage measurements to estimate $\nabla{h}$. Though we design and analyze the algorithm under a fixed-point condition,  the online implementation could asymptotically mitigate the model errors due to the closed-loop nature. This might shed light on why the online implementation can achieve a better performance than the offline implementation.
\section{Case Study}\label{sec:CaseStudy}
\subsection{Overview}
In this section, numerical simulations are performed in the modified IEEE 123-bus test case \cite{testfeeder} to demonstrate the proposed online voltage control scheme. As shown in Fig.~\ref{fig:IEEE123}, PV generators are distributed across the radial distribution network. 
{Note that the proposed online PNM-based VVC strategy can be embedded into the two-layer VVC framework to coordinate the conventional discrete voltage regulation devices and DERs in different timescales. In the upper layer, the conventional discrete voltage regulation devices are scheduled  over a slow timescale. In the lower layer, the VAr outputs of DERs can be adjusted by the online PNM-based VVC over a fast timescale. More details regarding the determination of  the conventional devices in the upper layer are given in Appendix B.}\footnote{{The  operating status of on-line load tap changer and capacitor banks are determined in the upper layer. }}
In the numerical simulations, the base voltage for the modified IEEE 123-bus network is 4.16 kV and the base power is 100 kVA. We set the reference of squared voltage magnitude as $\bm{v}_r=\bm{1}_m$, a $m\times1$ column vector of ones. With respect to the PNM algorithm, the parameter settings are: $\bm{C}$ is the identity matrix, $\varepsilon=0.001$, $\beta=0.5$, $\delta=0.1$.

 \begin{table}[t]
\centering
\caption{Objective Values and Convergence Iterations Computed by Different VVC Strategies.}\label{opt_compare}
\footnotesize
\renewcommand\arraystretch{1.0}
        \begin{tabular}{cccc}
        \hline
        \hline
         Strategy&Feedback&Convergence Iteration&Value \\
         \hline
         GP-based VVC&Yes&46& 0.0013\\
         DSGP-based VVC&Yes&25& 0.0007\\
         PNM-based VVC&Yes&5& 0.0005\\
         Offline OPF&No&\text{{converged}}& 0.0015\\
        \hline
        \hline
        \end{tabular}
\end{table}
 \begin{figure}[t]
     \centering
     \includegraphics[width=3.2in]{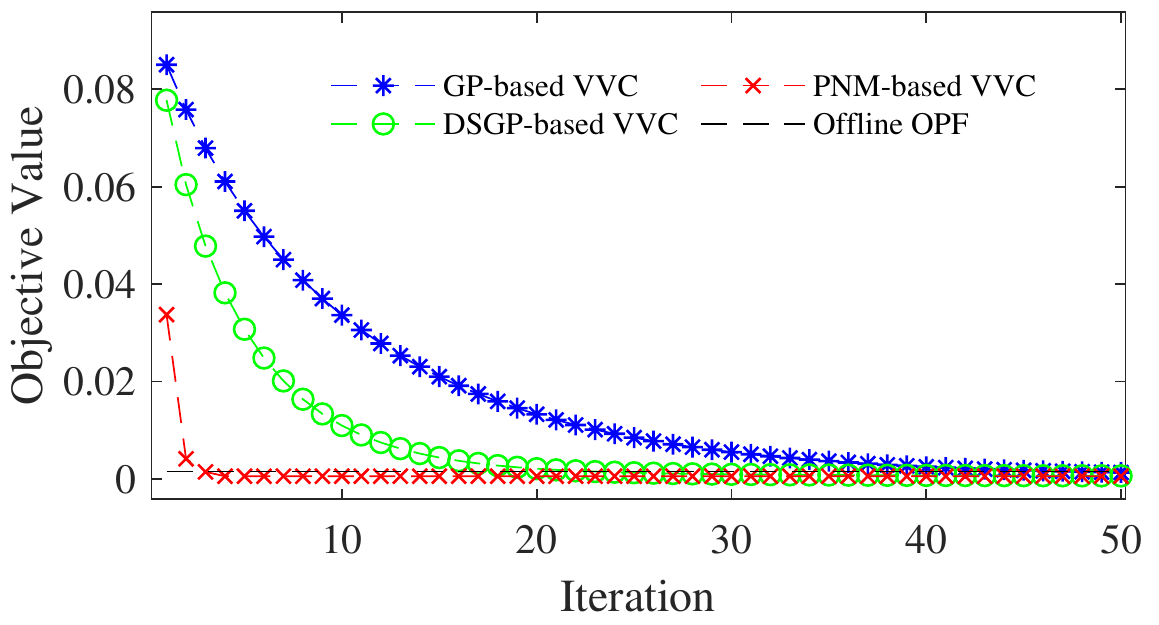}
     \caption{{Convergence performance of different VVC strategies.}}
     \label{fig:StaticCompare}
 \end{figure}
 
 To validate the effectiveness and superiority of the proposed online VVC strategy, both the static scenario and the dynamic scenario are taken into account in our simulations. The static scenario considers the static load and PV generators while the dynamic scenario considers the time-varying load and PV generators. All the simulations are conducted using MATLAB R2019b, the open source simulator OpenDSS \cite{OpenDss} and the IBM ILOG CPLEX 12.9 solver \cite{cplex}. 
\subsection{Static Scenario}
In the static scenario, each phase node has a constant load of  (6+j3) kVA, and each phase of PV generators supplies 20 kW real power generation to the grid. We also assume each phase of PV generator can
supply or consume at most 50 kVar reactive power. For the static scenario, we consider the following four different VVC strategies:
 
{(i) GP-based VVC (\cite{YG}): As shown in (\ref{eq:GP}), the GP-based VVC replies on the gradient information and projection operation to adjust the VAr outputs of DERs.} 

{(ii) DSGP-based VVC (\cite{HZ},\cite{WJ}): In the DSGP-based VVC, the diagonal entries of the Hessian matrix are leveraged to scale the gradient, thus improving the convergence performance. It adopts the scaled gradient and projection operation to adjust the VAr outputs of DERs.
}  

{(iii) PNM-based VVC: As shown in Algorithm 2, it leverages the projected Newton method to update the VAr outputs of DERs in the online manner.}

{(iv) Offline OPF: The problem (\ref{eq:Reformulation}) is directly solved by the CPLEX solver \cite{cplex} without considering the feedback of voltage measurements.}
 
 \begin{figure}[t]
     \centering
     \includegraphics[width=3in]{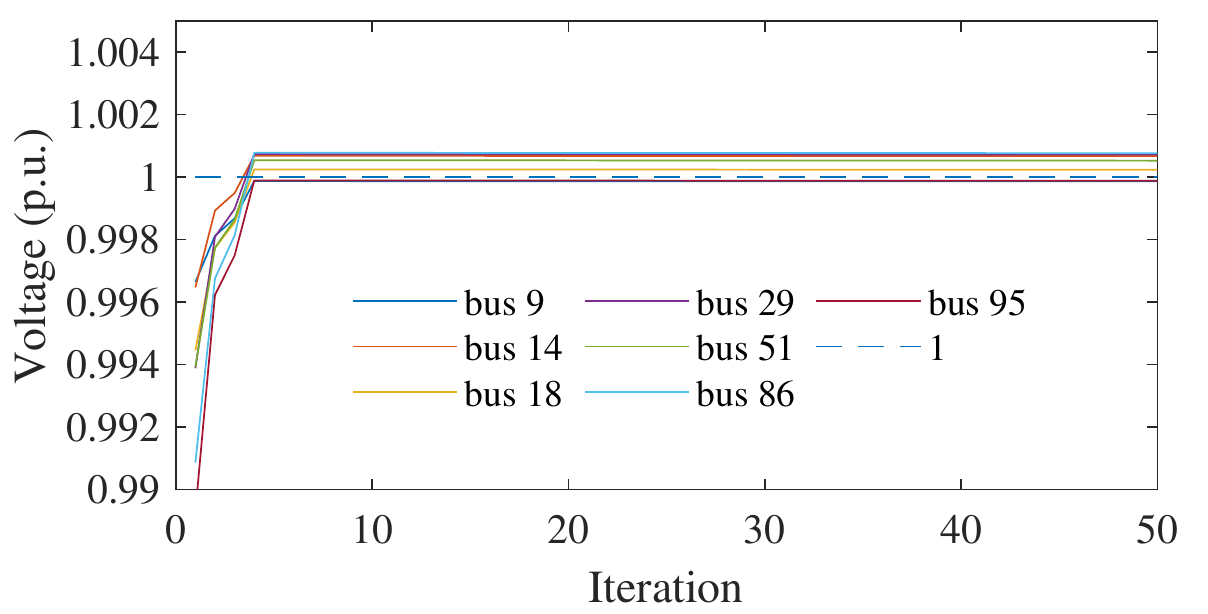}
     \caption{Voltage magnitudes of buses 9, 14, 18, 29, 51, 86, 95 in the phase-a network for the online PNM-based VVC under the static scenario.}
     \label{fig:StaticV}
 \end{figure}
  \begin{figure}[t]
     \centering
     \includegraphics[width=3in]{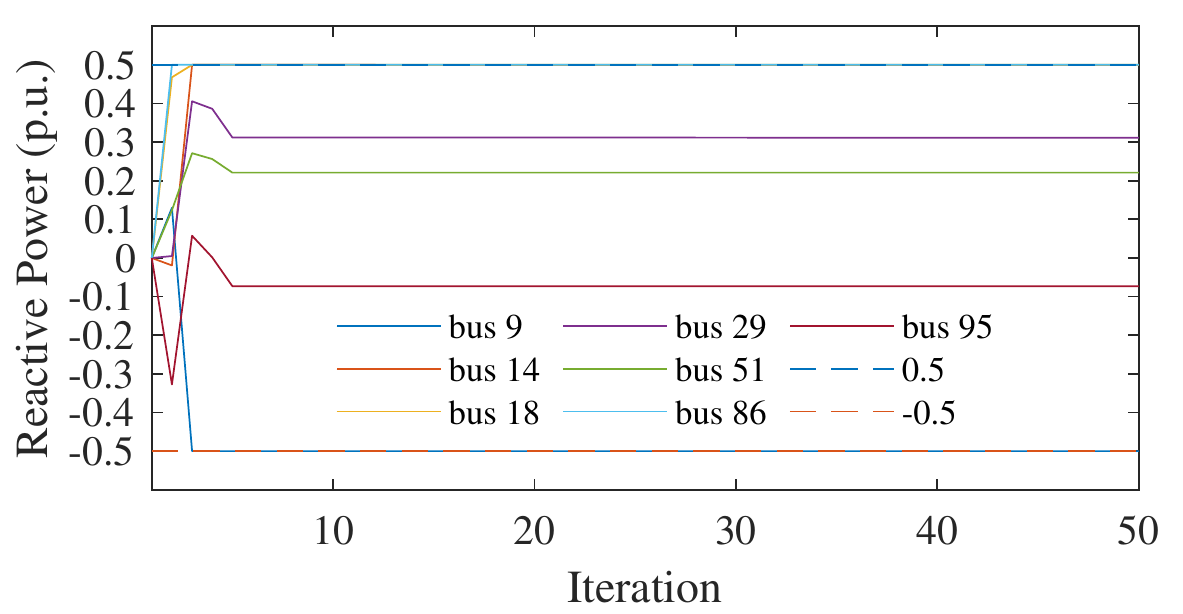}
     \caption{VAr outputs of buses 9, 14, 18, 29, 51, 86, 95 in the phase-a network for the online PNM-based VVC under the static scenario.}
     \label{fig:StaticReactivePower}
 \end{figure}
 \begin{figure}[b]
     \centering
     \includegraphics[width=3in]{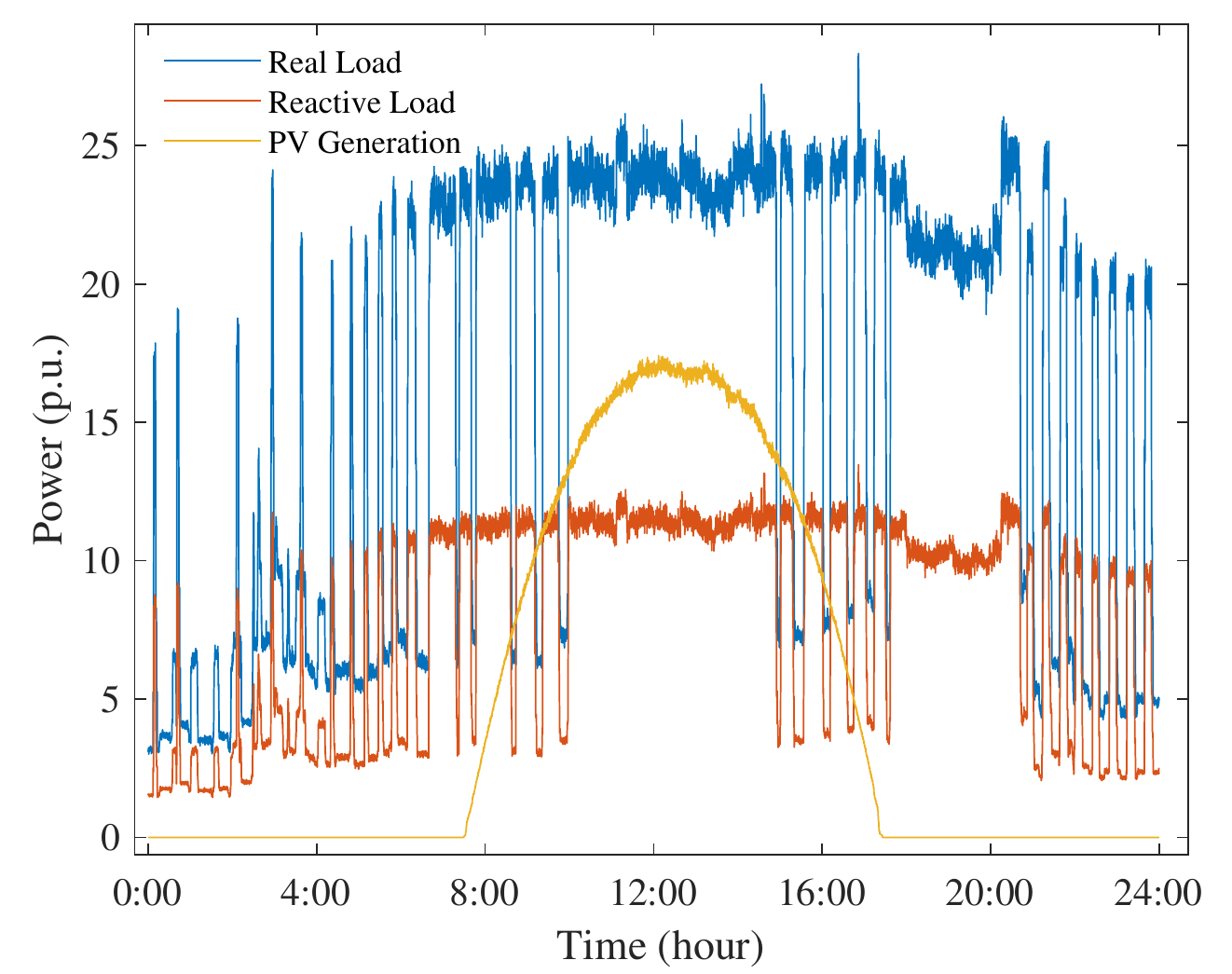}
     \caption{Aggregate load and PV generation across the IEEE 123-bus test system.}
     \label{fig:Aggregate_LoadPV}
\end{figure}
Note that voltage magnitudes, solved by the nonlinear distribution power flow in OpenDSS, are regarded as the actual voltage measurements to update the VAr outputs of PV generators in the GP-based VVC, DSGP-based VVC, and PNM-based VVC. Table II and Fig.\ref{fig:StaticCompare} show objective values and convergence iterations computed by different VVC strategies. It can be observed that the GP-based VVC, DSGP-based VVC and PNM-based VVC have smaller objective values than the OPF method due to the fact that they are able to asymptotically compensate the model errors via the feedback mechanism. The GP-based VVC shows a slow convergence rate since the convergence performance of the convention gradient-based method (steepest gradient) is not good. The performance of DSGP-based VVC is better than the GP-based VVC by taking advantage of the diagonal scaling. The convergence iteration of PNM-based VVC is only 5, which is far less than GP-based VVC and DSGP-based VVC, validating our previous analysis. This implies the online PNM-based VVC will have less communication costs in real-life implementations. Fig.\ref{fig:StaticV} and Fig.\ref{fig:StaticReactivePower} show that voltage magnitudes and VAr outputs of buses 9, 14, 18, 29, 51, 86, 95 in the phase-a network as the PNM-based VVC is applied. As depicted in Fig.\ref{fig:StaticV} and Fig.\ref{fig:StaticReactivePower}, the voltage magnitudes are close to the target voltage 1.0 without violating the VAr limits by utilizing the PNM-based VVC.
\subsection{Dynamic Scenario}
For the dynamic scenario,  we consider a more realistic time-varying system. The aggregate load and PV generation across the modified IEEE 123-bus test case  are shown in Fig.\ref{fig:Aggregate_LoadPV}, where the total time span is one day (24 hours) and the time resolution is 10s. 
\begin{figure}[t]
     \centering
     \includegraphics[width=3in]{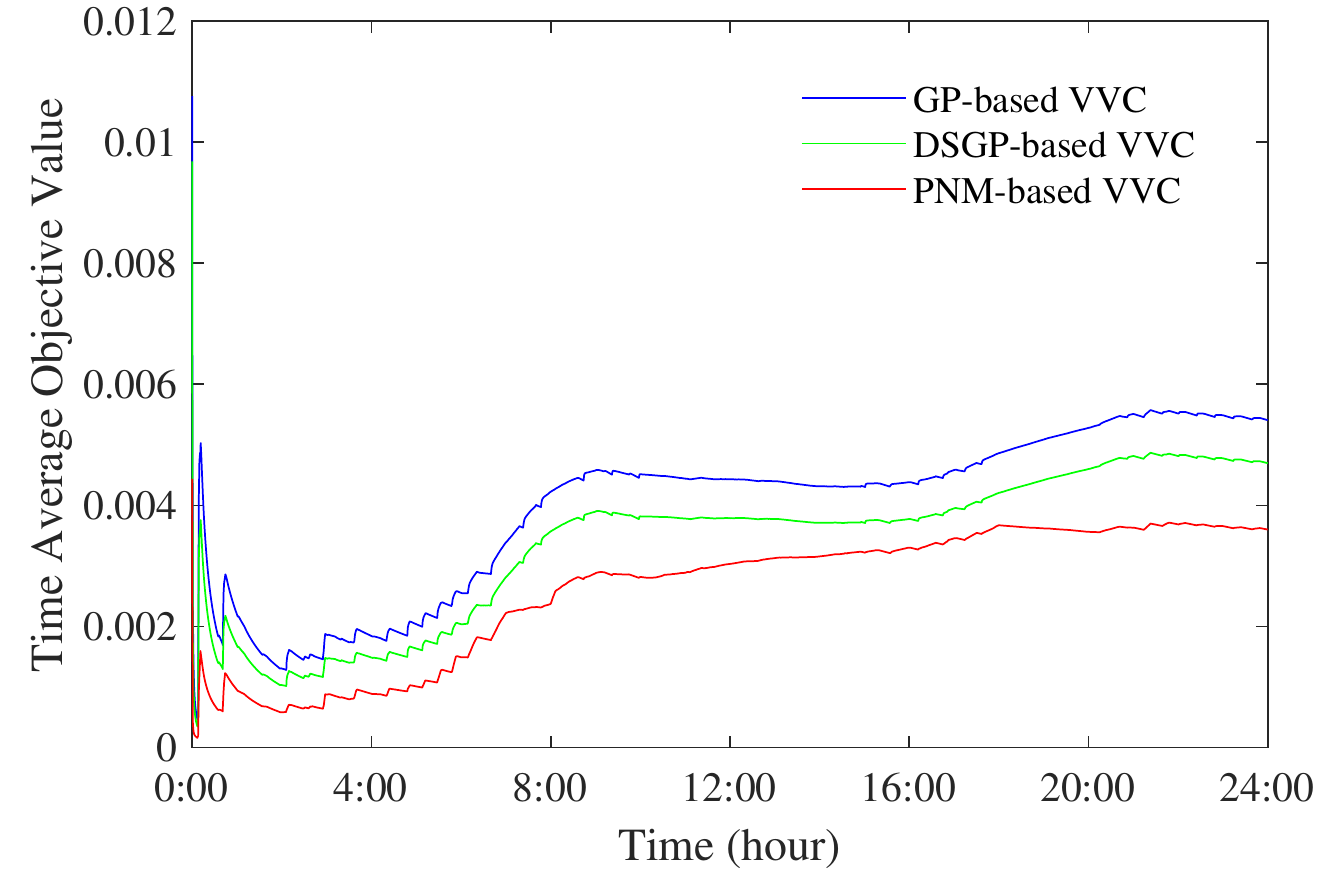}
     \caption{{Time average objective values with different VVC strategies under the dynamic scenario.}}
     \label{fig:DynamicCompare}
 \end{figure}
 
 \begin{figure}[t]
     \centering
     \includegraphics[width=3in]{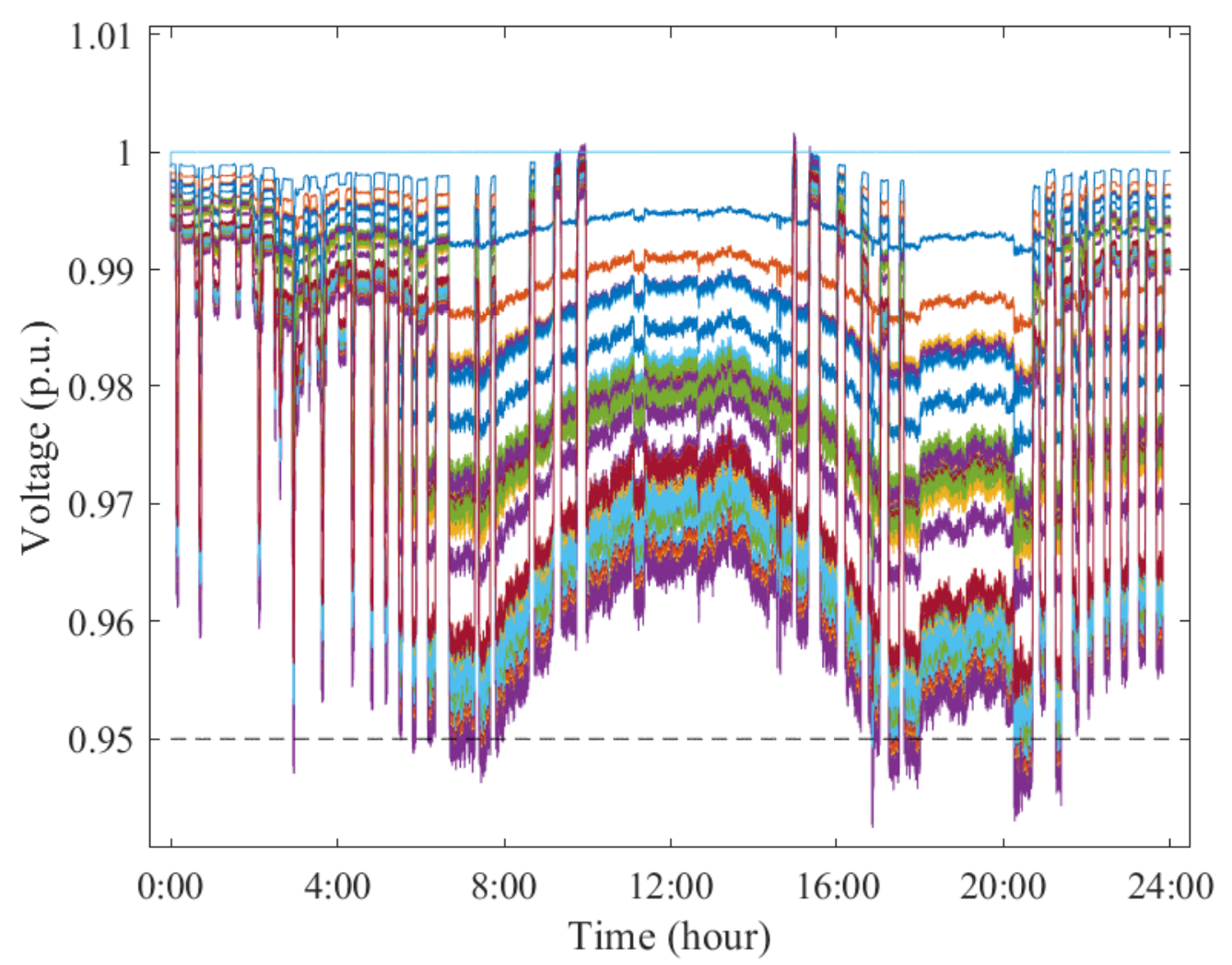}
     \caption{Voltage profiles without control in the phase-a network (each curve depicts the voltage magnitude fluctuation for each phase-a node of bus) under the dynamic scenario.}
     \label{fig:Va_WithoutControl}
\end{figure}
   \begin{figure}[t]
     \centering
     \includegraphics[width=3in]{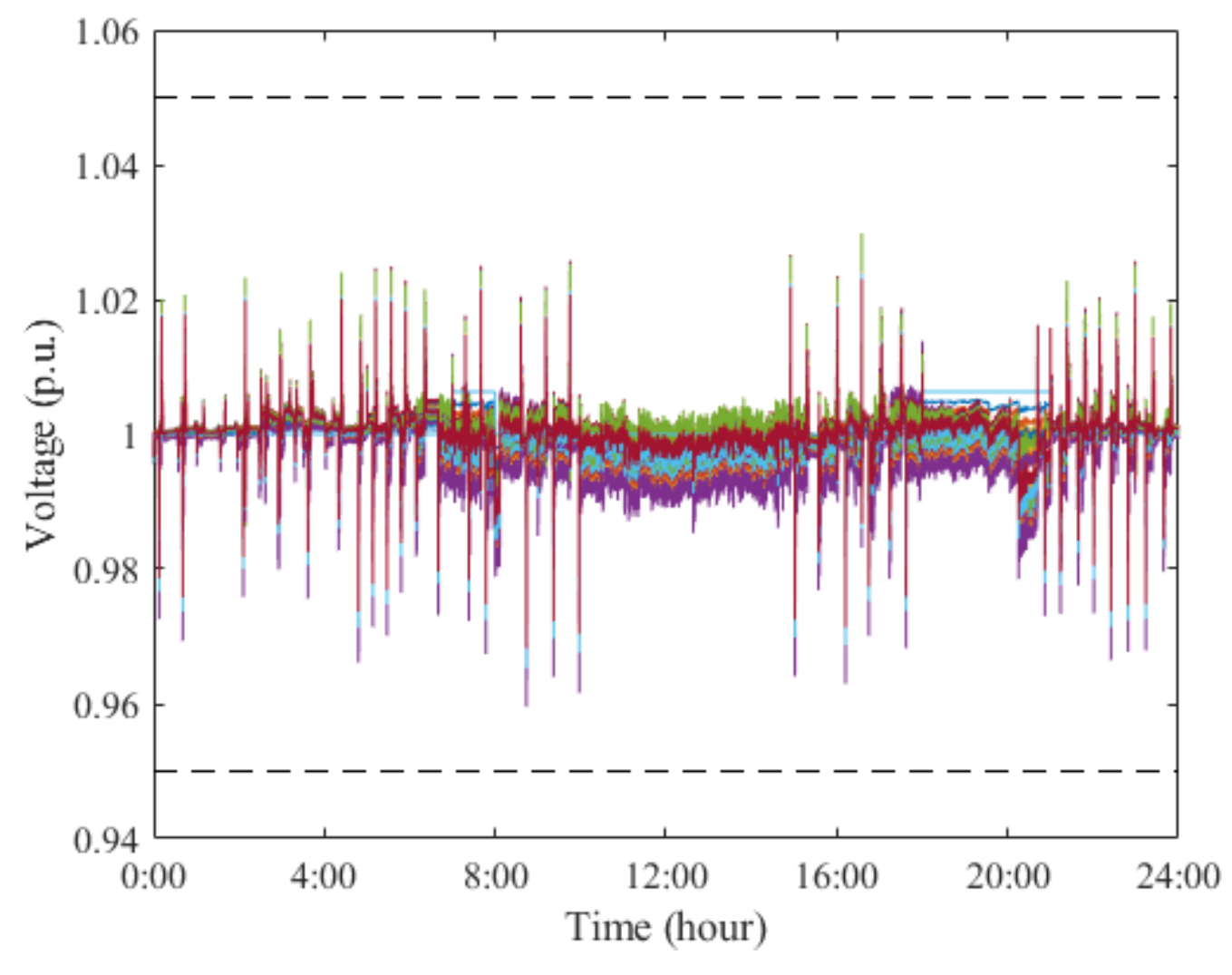}
     \caption{{Voltage profiles with the online PNM-based VVC in the phase-a network (each curve depicts the voltage magnitude fluctuation for each phase-a node of bus) under the dynamic scenario.}}
     \label{fig:Va_WithControl}
 \end{figure}
   \begin{figure}[t]
     \centering
     \includegraphics[width=3in]{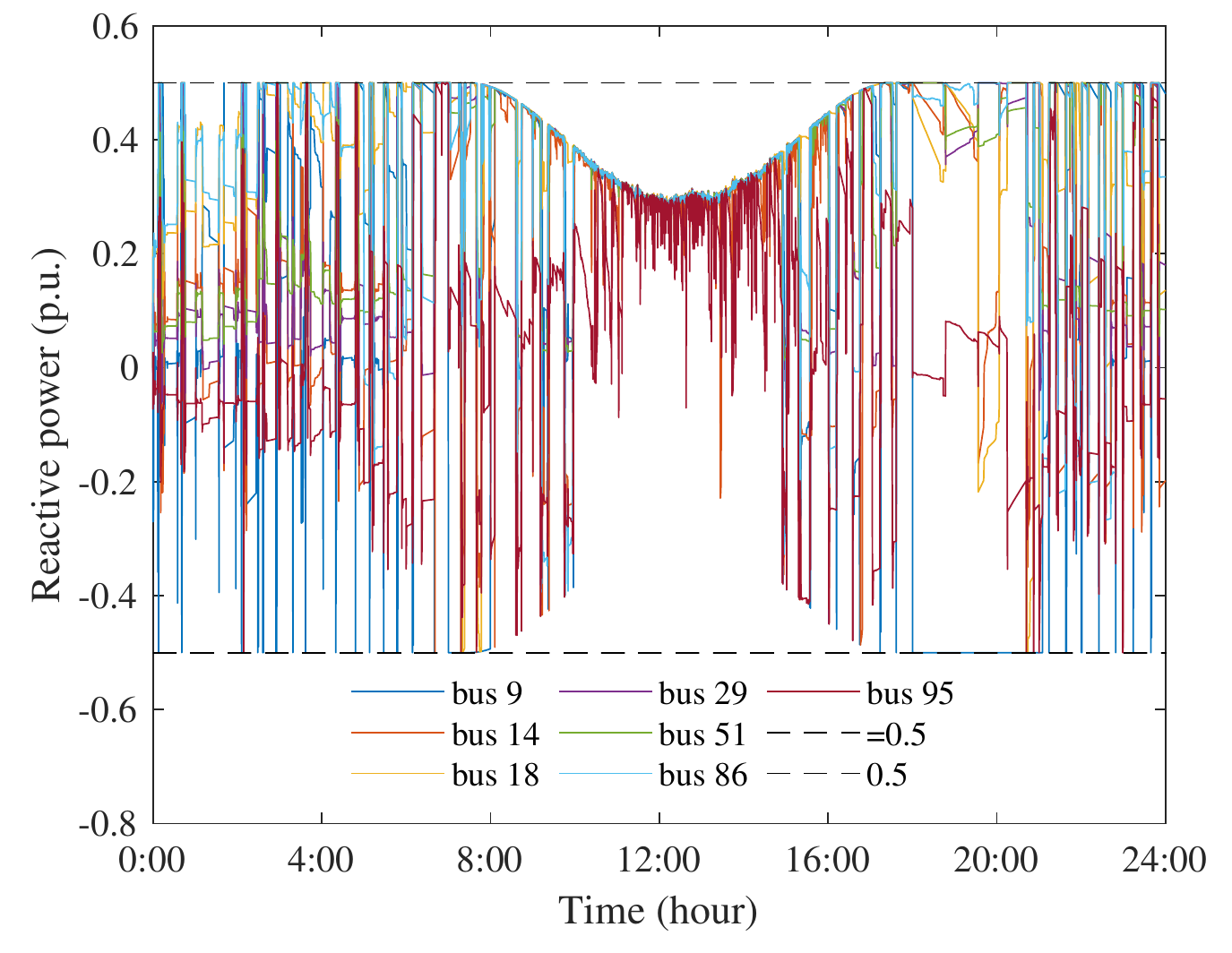}
     \caption{{VAr outputs of buses 9, 14, 18, 29, 51, 86, 95 in the phase-a network for the online PNM-based VVC under the dynamic scenario.}}
     \label{fig:DynamicVAr}
 \end{figure}

In the dynamic scenario, three different VVC strategies, including GP-based, DSGP-based, PNM-based VVC, are taken into account\footnote{Considering the offline OPF can only be implemented after convergence, it might not be suitable to apply the offline OPF to the dynamic scenario with a fast change rate. Thus, it is not carried out in the dynamic scenario.}. The control period of those VVC strategies is set as 2s, i.e., the VAr outputs of PV inverters are updated every 2 seconds. In the dynamic scenario, the capacities of PV inverters are set as 50 kVA. And the VAr limits $\bm{\underline{q}}^g,\bm{\overline{q}}^g$ are updated online based on the given inverter capacities and the instantaneous real power of PV generators. Time average objective values across time steps are chose as the index to test the performance of those different VVC strategies in the dynamic scenario. Fig.\ref{fig:DynamicCompare} shows the time average objective values across time steps computed by different VVC strategies. It can be observed from Fig.\ref{fig:DynamicCompare} that the PNM-based VVC exhibits a better performance compared with GP-based VVC and DSGP-based VVC. Due to the fast convergence of PNM, the online PNM-based VVC has a great tracking capability to follow the time-varying changes in the system, thus leading to the better performance of PNM-based VVC compared with GP-based and DSGP-based VVC.

Taking phase a as an example, Fig.\ref{fig:Va_WithoutControl} and Fig.\ref{fig:Va_WithControl} show voltage profiles without control and with the online PNM-based VVC in the phase-a network, respectively. In addition, Fig.\ref{fig:DynamicVAr}
exhibits the VAr outputs of buses 9, 14,18, 29, 51, 86, 95 under the dynamic scenario.
It can be seen that there are under-voltage violations (below 0.95 p.u.) around 8:00, 16:00 and 20:00 for the distribution network without control due to the high load demands but low PV generation around those periods. In contrast, there is no any voltage violation across one day as the online PNM-based VVC is applied to the distribution network. It means  the proposed online PNM-based VVC  can effectively eliminate voltage violations in the dynamic scenario. 

\subsection{{Algorithm Scalability: IEEE 8500-node Test Feeder}}

\begin{figure}[t]
    \centering
    \includegraphics[width=3in]{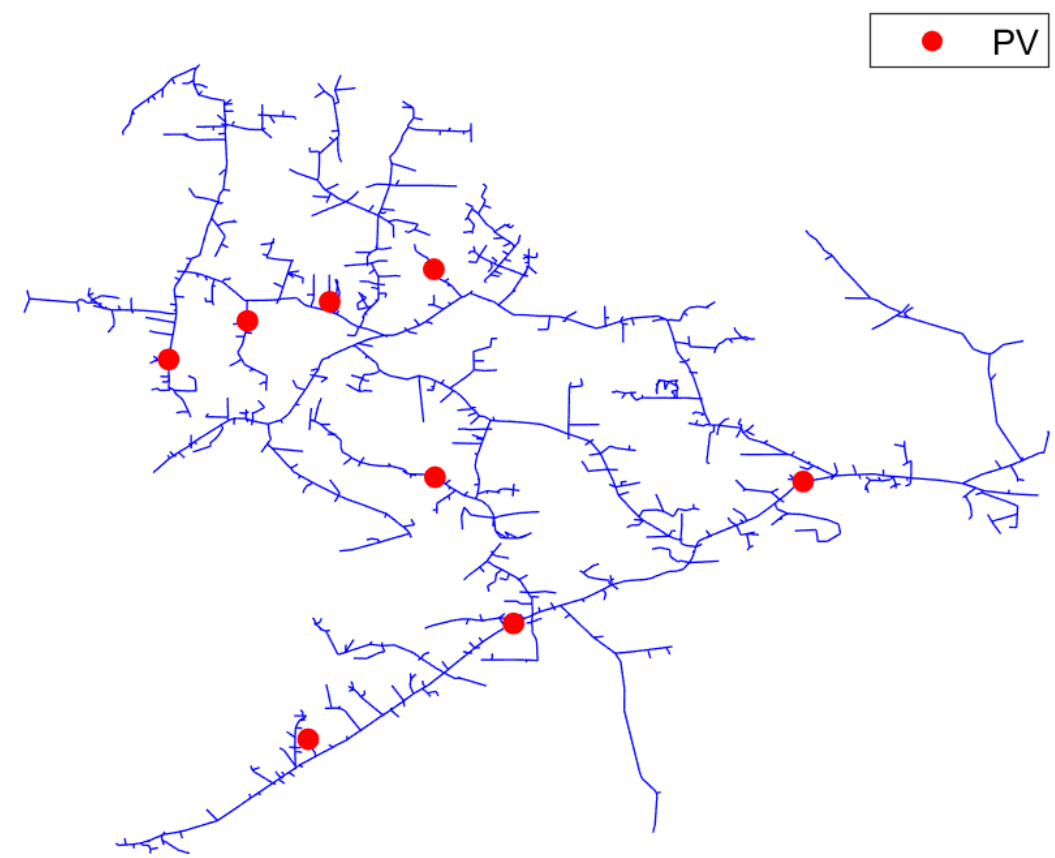}
    \caption{{IEEE 8500-node test feeder.}}
    \label{fig:8500}
\end{figure}

\begin{figure}[t]
    \centering
    \includegraphics[width=3in]{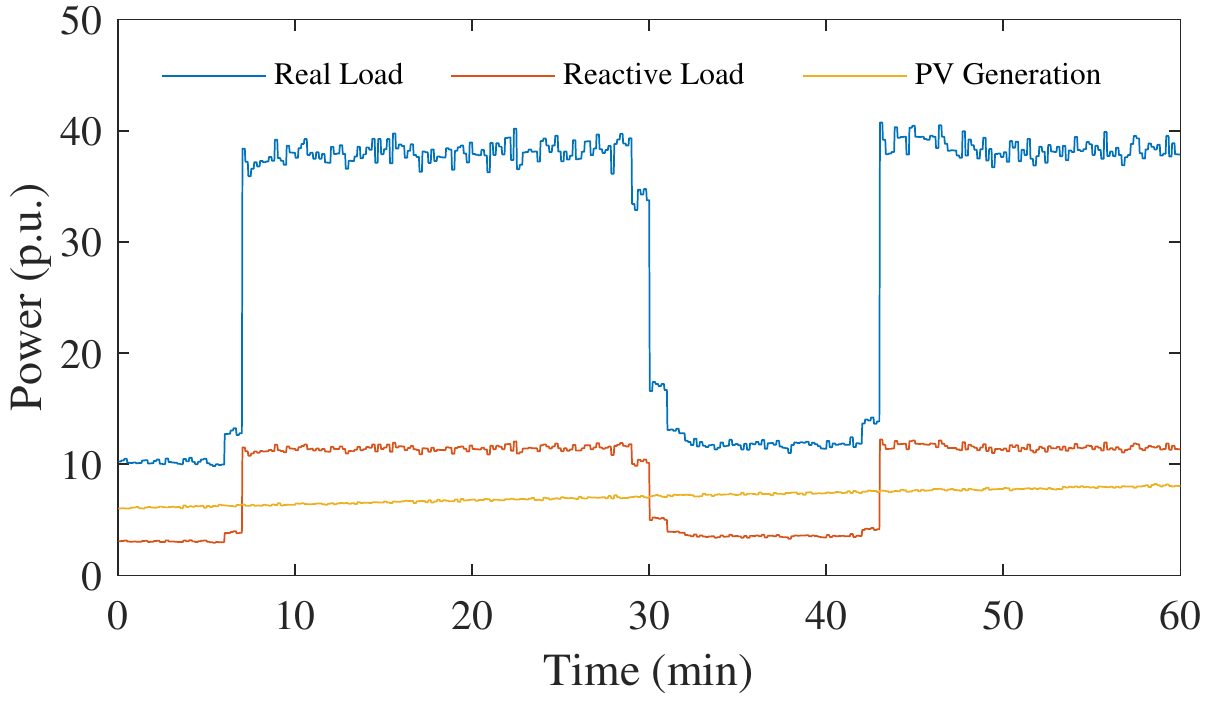}
    \caption{{Aggregate load and PV generation across the IEEE 8500-node test feeder.}}
    \label{fig:8500load}
\end{figure}

{The IEEE 8500-node test feeder \cite{8500} is used to test the scalability of the online PNM-based VVC strategy. The test feeder is modified by adding several PV generators, as shown in Fig.\ref{fig:8500}, where the capacities of thoes PV inverters are set as 100 kVA. The base voltage for the IEEE 8500-node test feeder is 7.2 kV and the base power is 100 kVA. The aggregate load and PV generation across the IEEE 8500-node test feeder are shown in Fig.\ref{fig:8500load}. 
}

\begin{figure}[t]
    \centering
    \includegraphics[width=3in]{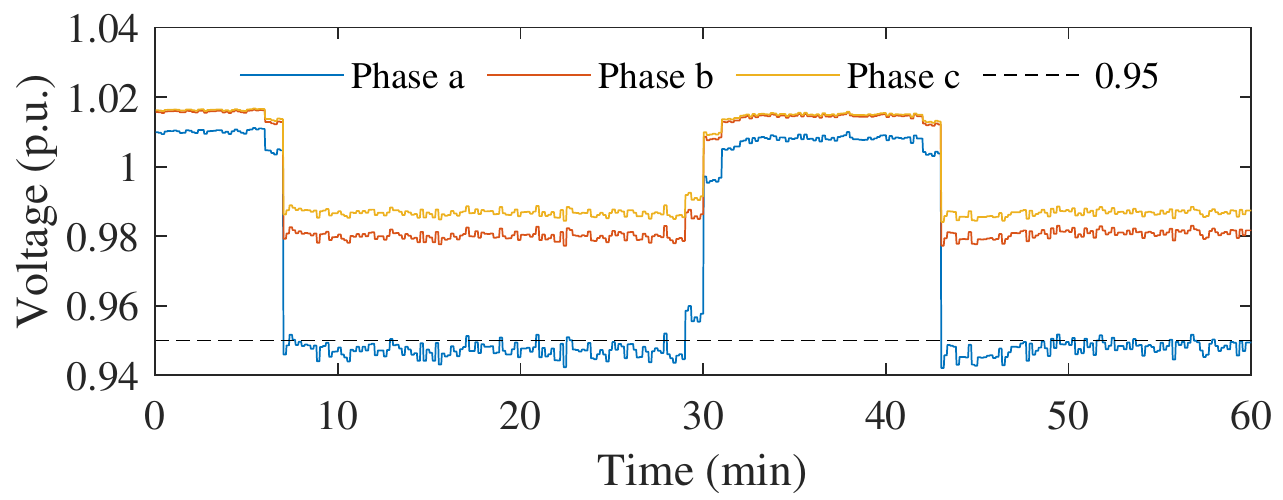}
    \caption{{Minimum voltage profiles without control in phase-a, phase-b and phase-c networks for each time step.}}
    \label{fig:8500VoltWithoutControl}
\end{figure}

\begin{figure}[t]
    \centering
    \includegraphics[width=3in]{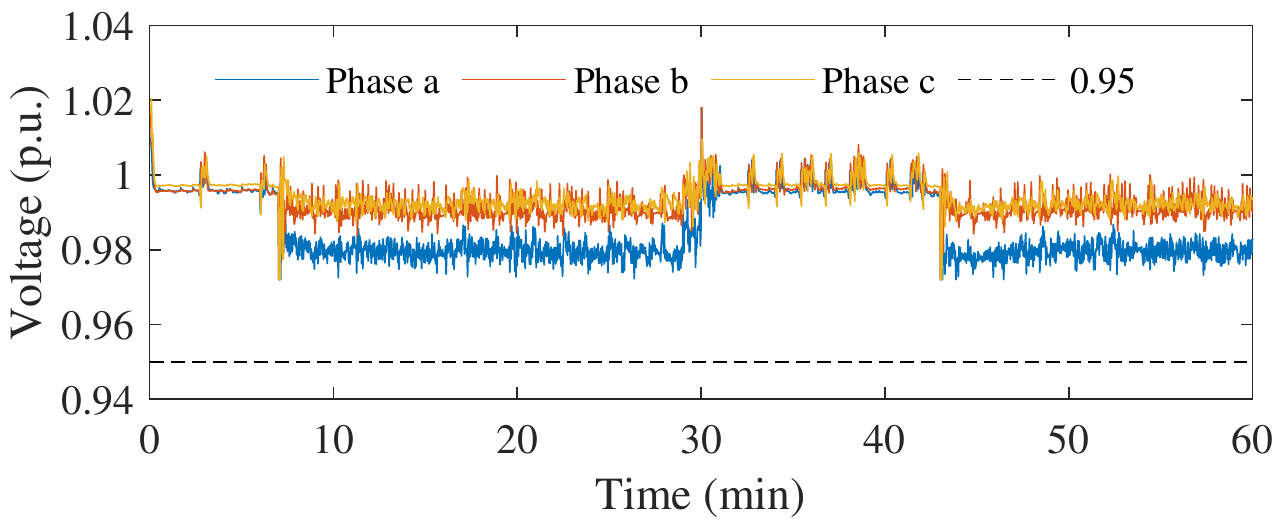}
    \caption{{Minimum voltage profiles with the online PNM-based VVC in phase-a, phase-b and phase-c networks for each time step.}}
    \label{fig:8500VoltWithPNM}
\end{figure}

{
Fig. \ref{fig:8500VoltWithoutControl} reports minimum voltage profiles without control in phase-a, phase-b and phase-c networks for each time step. As seen in Fig. \ref{fig:8500VoltWithoutControl} , voltage violations occur in phase-a networks for this distribution network without control. Next, suppose one change is made: the proposed online PNM-based VVC is applied to manage the VAr outputs of PV generators. In this case, the maximum voltage magnitude across this IEEE 8500-node test feeder over time steps is 1.036  when implementing the online PNM-based VVC. In addition, Fig. \ref{fig:8500VoltWithPNM}  reports minimum voltage profiles with the online PNM-based VVC in phase-a, phase-b and phase-c networks for each time step. As seen in Fig. \ref{fig:8500VoltWithPNM}, no voltage violation occurs in phase-a, phase-b, and phase-c networks. These results illustrate the online PNM-based VVC is still effective to protect against voltage violations for this large distribution network, validating the algorithm scalability of our proposed online PNM-based VVC. 
}

\section{Conclusion}\label{sec:Conclusion}
This paper proposes an online voltage control method for unbalanced distribution networks where PNM is leveraged to improve the convergence performance. Through the feedback of voltage measurements from local bus agents, the central agent can effectively coordinate the VAr outputs of DERs in real time to cope with  fast  system  variations. The theoretical analysis and proof of this work are established on unbalanced distribution networks.

Numerical case studies are conducted in both static and dynamic scenarios to test the performance of the proposed online PNM-based VVC. Through numerical case studies, {it is} found that the online voltage control with feedback exhibits a better performance than offline OPF without feedback since the closed-loop nature of online feedback can asymptotically compensate the model errors via the feedback mechanism. Our proposed online PNM-based VVC can effectively eliminate the voltage violations in unbalanced distribution networks, and shows a better control performance compared with other methods (e.g., GP and DSGP), validating our theoretical analysis.

{
This work is the model-based VVC strategy, relying on the knowledge of distribution networks. However, it is promising to reduce the requirement for the knowledge of distribution networks due to the increasing deployment of advanced monitoring and metering infrastructures and available data in the distribution network level. Our future work will investigate data-driven model-free VVC strategies based on advanced machine learning techniques.
}

\section*{Appendix A}
\medskip
\noindent
\textbf{Proof of Proposition 4:} Making use of \cite[Prop.~2.1.2]{BertsekasNP}, it follows that $\bm{q}^{g\ast}$ is the optimal solution of (\ref{eq:Reformulation}), if and only if:
\begin{equation}\label{eq:FirstOrder}
    (\bm{q}^g(t)-\bm{q}^{g\ast})^T\nabla{h}(\bm{q}^{g\ast})\geq{0}, \forall{\bm{q}^g(t)}\in[\bm{\underline{q}}^g,\bm{\overline{q}}^g]
\end{equation}
Now, we show the sufficient and necessary conditions of (\ref{eq:FirstOrder}) are:
\begin{subequations}\label{eq:OptimalityCondition}
\begin{align}
    &\frac{\partial{h}(\bm{q}^{g\ast})}{\partial[\bm{q}^g]_i}\geq{0} \text{~if~} [\bm{q}^{g\ast}]_i=[\bm{\underline{q}}^g]_{i}\\
    &\frac{\partial{h}(\bm{q}^{g\ast})}{\partial[\bm{q}^g]_i}\leq{0} \text{~if~} [\bm{q}^{g\ast}]_i=[\bm{\overline{q}}^g]_{i}\\
    &\frac{\partial{h}(\bm{q}^{g\ast})}{\partial[\bm{q}^g]_i}=0,
    \text{~if~} [\bm{\underline{q}}^g]_{i}<[\bm{q}^{g\ast}]_i<[\bm{\overline{q}}^g]_{i}
\end{align}
\end{subequations}

\noindent
\textbf{Sufficiency:} As the condition (\ref{eq:OptimalityCondition}) holds, it is straightforward that (\ref{eq:FirstOrder}) holds.

\noindent
\textbf{Necessity:} It will be established using proof by contrapositive. Suppose there exists a $[\bm{q}^{g\ast}]_i$ such that the condition (\ref{eq:OptimalityCondition}) fails to hold, there always exists a $\bm{q}^g(t)\in[\bm{\underline{q}}^g,\bm{\overline{q}}^g]$ such that the condition (\ref{eq:FirstOrder}) fails to hold. For example, assume (\ref{eq:OptimalityCondition}a) does not hold, i.e., $\frac{\partial{h}(\bm{q}^{g\ast})}{\partial[\bm{q}^g]_i}<0 \text{~as~} [\bm{q}^{g\ast}]_i=[\bm{\underline{q}}^g]_{i}$. Then, let all the elements in $\bm{q}^g(t)$ be the same as $\bm{q}^{g\ast}$ except $[\bm{q}^{g}(t)]_i>[\bm{\underline{q}}^{g}]_{i}=[\bm{q}^{g\ast}]_i$. In this case, such a $\bm{q}^g(t)$ does not satisfy  (\ref{eq:FirstOrder}). Thus, we conclude that (\ref{eq:OptimalityCondition}) is the necessary condition of (\ref{eq:FirstOrder}). 

From the above analysis, we can know that if $\bm{q}^g(t)$ is the optimal solution $\bm{q}^{g\ast}$, then (\ref{eq:OptimalityCondition})  holds. It follows from (\ref{eq:OptimalityCondition}) that [P4.a] holds.

We then prove [P4.b] holds. Note that with respect to $i\in\bm{I}(t)$, it follows from the definition of $\bm{I}(t)$ and Proposition 3 that:
\begin{subequations}\label{eq:UiInIK}
\begin{align}
    &[\bm{u}(t)]_i>0 \mbox{~if~}[\bm{\underline{q}}^g]_{i}\leq[\bm{q}^g(t)]_{i}\leq[\bm{\underline{q}}^g]_{i}+[\bm{\varepsilon}(t)]_i, i\in\bm{I}(t)\\
    &[\bm{u}(t)]_i<0 \mbox{~if~}[\bm{\overline{q}}^g]_{i}-[\bm{\varepsilon}(t)]_i\leq[\bm{q}^g(t)]_i\leq[\bm{\overline{q}}^g]_{i}, i\in\bm{I}(t)
\end{align}
\end{subequations}
Consider the following sets of indices:
\begin{subequations}\label{eq:ISet}\small
\begin{align}
    \nonumber\bm{I}_1=&\{i|[\bm{\overline{q}}^g]_i>[\bm{q}^g(t)]_i>[\bm{\underline{q}}^g]_i\mbox{~or~}[\bm{q}^g(t)]_i=[\bm{\underline{q}}^g]_i, [\bm{u}(t)]_i<0\\
    &\mbox{or~}[\bm{q}^g(t)]_i=[\bm{\overline{q}}^g]_i, [\bm{u}(t)]_i>0 \mbox{~for~}
    i\in\{1,..,m\}\text{\textbackslash}\bm{I}(t)
    \}\\
     \nonumber\bm{I}_2=&\{i|[\bm{q}^g(t)]_i=[\bm{\underline{q}}^g]_i, [\bm{u}(t)]_i\geq{0} \mbox{~or~}[\bm{q}^g(t)]_i=[\bm{\overline{q}}^g]_i, \\&[\bm{u}(t)]_i\leq{0}\mbox{~for~} i\in\{1,..,m\}\text{\textbackslash}\bm{I}(t)
     \}\\
     \bm{I}_3=&\{i|[\bm{q}^g(t)]_i=[\bm{\underline{q}}^g]_i\mbox{~or~}[\bm{q}^g(t)]_i=[\bm{\overline{q}}^g]_i, \mbox{~for~} i\in\bm{I}(t)\}\\
     \nonumber\bm{I}_4=&\{i|[\bm{\underline{q}}^g]_i<[\bm{q}^g(t)]_i\leq[\bm{\underline{q}}^g]_i+[\bm{\varepsilon}(t)]_i\\
     &\mbox{~or~}{[\bm{\overline{q}}^g]_i-[\bm{\varepsilon}}(t)]_i{\leq}[\bm{q}^g(t)]_i<[\bm{\overline{q}}^g]_i, \mbox{~for~} i\in\bm{I}(t)\}
\end{align}
\end{subequations}
And set $\tilde{\alpha}$:
\begin{equation}\small
    \tilde{\alpha}=\sup\{\alpha(k)|~[\bm{\overline{q}}^g]_i\geq[\bm{q}^g(t)]_i-\alpha(t)[\bm{u}(t)]_i\geq[\bm{\underline{q}}^g]_i,~\forall{i}\in\bm{I}_1\cup\bm{I}_4\}
\end{equation}
In view of the definitions of $\bm{I}_1$ and $\bm{I}_4$ as well as (\ref{eq:UiInIK}), we can know $\tilde{\alpha}>0$. Define the vector $\bar{\bm{u}}(t)$ as follows:
\begin{equation}\label{eq:BarU}
[\bm{\bar{u}}(t)]_i=
    \begin{cases}
     [\bm{u}(t)]_i, i\in\bm{I}_1\cup\bm{I}_4\\
     0, i\in\bm{I}_2\cup\bm{I}_3
    \end{cases}
\end{equation}
In view of (\ref{eq:ScaledGradient}) and  (\ref{eq:UiInIK})-(\ref{eq:BarU}), we have:
\begin{equation}\label{eq:TildeAlpha}
\begin{split}
    \bm{q}^g(t+1)&=[\bm{q}^g(t)-\alpha(t)\bm{u}(t)]_{\bm{\underline{q}}^g}^{\bm{\overline{q}}^g}\\&=\bm{q}^g(t)-\alpha(t)\bm{\bar{u}}(t), \forall{\alpha(t)}\in(0,\tilde{\alpha}]
\end{split}
\end{equation}
In view of the definitions of $\bm{I}_2$ and $\bm{I}(t)$, it follows that:
\begin{subequations}\label{eq:I2}
\begin{align}
    \frac{\partial{h}(\bm{q}^g(t))}{\partial{[\bm{q}^g]_i}}\leq{0}, &\mbox{~if~}i\in\bm{I}_2,~[\bm{q}^g(t)]_i=[\bm{\underline{q}}^g]_i\\
    \frac{\partial{h}(\bm{q}^g(t))}{\partial{[\bm{q}^g]_i}}\geq{0}, &\mbox{~if~}i\in\bm{I}_2,~[\bm{q}^g(t)]_i=[\bm{\overline{q}}^g]_i
\end{align}
\end{subequations}
Combining (\ref{eq:ISet}b) and (\ref{eq:I2}), it follows that: 
\begin{equation}\label{eq:ExtensionofI2}
    \sum_{i\in\bm{I}_2}\frac{\partial{h}(\bm{q}^g(t))}{\partial{[\bm{q}^g]_i}}[\bm{u}(t)]_i\leq{0}
\end{equation}
Using (\ref{eq:BarU}) and (\ref{eq:ExtensionofI2}), it follows that:
\begin{equation}\label{eq:AddI2}
\begin{split}
    &\nabla{h}(\bm{q}^g(t))^T\bar{\bm{u}}(t)=\sum_{i\in\bm{I}_1\cup\bm{I}_4}\frac{\partial{h}(\bm{q}^g(t))}{\partial{[\bm{q}^g]_i}}[\bm{u}(t)]_i\\&\geq{\sum_{i\in\bm{I}_1\cup\bm{I}_2}}\frac{\partial{h}(\bm{q}^g(t))}{\partial{[\bm{q}^g]_i}}[\bm{u}(t)]_i+\sum_{i\in\bm{I}_4}\frac{\partial{h}(\bm{q}^g(t))}{\partial{[\bm{q}^g]_i}}[\bm{u}(t)]_i
\end{split}
\end{equation}
Using (\ref{eq:U(K)}) and the structure of $\bm{D}(t)$, and rearranging terms, we have:
\begin{equation}\label{eq:SumI1I2I4}\small
\begin{split}
    &{\sum_{i\in\bm{I}_1\cup\bm{I}_2}}\frac{\partial{h}(\bm{q}^g(t))}{\partial{[\bm{q}^g]_i}}[\bm{u}(t)]_i+\sum_{i\in\bm{I}_4}\frac{\partial{h}(\bm{q}^g(t))}{\partial{[\bm{q}^g]_i}}[\bm{u}(t)]_i\\&=\bm{y}(t)^T\bm{B}(t)^{-1}\bm{y}(t)+\sum_{i\in\bm{I}_4}\big|[\bm{H}]_{ii}\big|^{-1}[\frac{\partial{h}(\bm{q}^g(t))}{\partial{[\bm{q}^g]_i}}]^2\geq{0}
\end{split}
\end{equation}
where $\bm{B}(t)$ is a positive definite matrix, defined in (\ref{eq:TildeE}), and $\bm{y}(t)$ is a column vector formed by rearranging $\frac{\partial{h}(\bm{q}^g(t))}{\partial{[\bm{q}^g]_i}}$ for ${i}\in{\bm{I}_1\cup\bm{I}_2}$ in the order consistent with $\bm{B}(t)$.

From the definitions of $\bm{I}_3$ and $\bm{I}(t)$, we can know that for $i\in\bm{I}_3$, $\frac{\partial{h}(\bm{q}^g(t))}{\partial{[\bm{q}^g]_i}}$ satisfies (\ref{eq:OptimalityCondition}).  We next prove: as $\bm{q}^g(t)$ is not the optimal solution, there must exist  a $\frac{\partial{h}(\bm{q}^g(t))}{\partial{[\bm{q}^g]_i}}\neq{0}$ for $i\in\{\bm{I}_1\cup\bm{I}_2\cup\bm{I}_4\}$, which can be proved by contradiction. Suppose  $\frac{\partial{h}(\bm{q}^g(t))}{\partial{[\bm{q}^g]_i}}={0}$ for ${i}\in\{\bm{I}_1\cup\bm{I}_2\cup\bm{I}_4\}$, then it follows that for $i\in\{\bm{I}_1\cup\bm{I}_2\cup\bm{I}_4\}$, $\frac{\partial{h}(\bm{q}^g(t))}{\partial{[\bm{q}^g]_i}}$ satisfies (\ref{eq:OptimalityCondition}). Plus that for $\forall{i}\in{\bm{I}_3}$, $\frac{\partial{h}(\bm{q}^g(t))}{\partial{[\bm{q}^g]_i}}$ satisfies (\ref{eq:OptimalityCondition}), it follows that 
$\bm{q}^g(t)$ is the optimal solution, resulting in the contradiction. Thus, we can know there must exist  a $\frac{\partial{h}(\bm{q}^g(t))}{\partial{[\bm{q}^g]_i}}\neq{0}$ for $i\in\{\bm{I}_1\cup\bm{I}_2\cup\bm{I}_4\}$. Combing it with (\ref{eq:SumI1I2I4}), it implies that:
\begin{equation}
    \bm{y}(t)^T\bm{B}(t)^{-1}\bm{y}(t)+\sum_{i\in\bm{I}_4}\big|[\bm{H}]_{ii}\big|^{-1}[\frac{\partial{h}(\bm{q}^g(t))}{\partial{[\bm{q}^g]_i}}]^2>0
\end{equation}
It then follows that:
\begin{equation}
    \nabla{h}(\bm{q}^g(t))^T\bar{\bm{u}}(t)>0
\end{equation}
Combing this relation with (\ref{eq:TildeAlpha}) and the fact $\tilde{\alpha}>0$, it follows that $\bar{\bm{u}}(t)$ is a feasible descent direction at $\bm{q}^g(t)$ and there exists a scalar $\bar{\alpha}(t)$ for which the desired relation [P4.~b] is satisfied. Q.E.D.

\section*{Appendix B}
\medskip
\noindent
{\textit{The determination of  the conventional devices in the upper layer }}

{The voltage control devices in distribution networks exhibit different temporal characteristics, which should be addressed in the voltage control system design. The conventional discrete devices with slow and discrete nature should be operated in the slow timescale. Frequent adjustments of those discrete devices are not beneficial for the economic operation. However, the inverter-based DERs, e.g., PV, with fast and continuous nature could effectively respond to voltage issues with a fast timescale. Consequently, those voltage regulation devices should be operated in different timescales.}

\begin{figure}[t]
     \centering
     \includegraphics[width=3.4in]{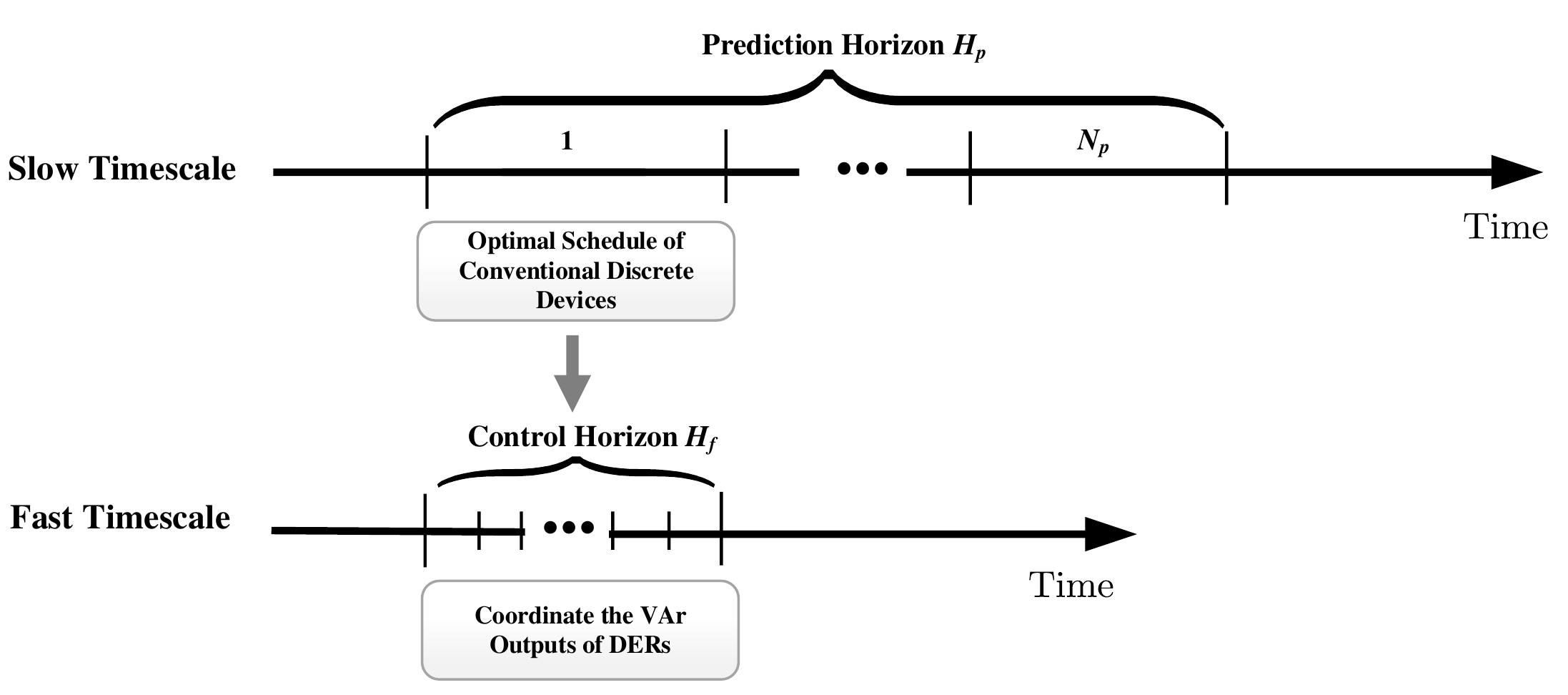}
     \caption{{Two-layer Volt/VAr control framework.}}
     \label{fig:Timescale}
 \end{figure}

{
The two-layer VVC framework is depicted in Fig.\ref{fig:Timescale}. In the upper layer, a model predictive control (MPC)-based control scheme is proposed to optimally coordinate the operation of conventional discrete devices, including on-load tap changer (OLTC) and capacitor banks (CBs), over slow timescale. Although the VAr outputs of DERs are also considered in the upper layer, only conventional discrete devices are scheduled in this upper layer.
}

{MPC is widely employed to make optimal operation decisions in over a given time horizon. Optimal operation decisions over this given time horizon are generated, but only the decision for the first time step is implemented in practice.} {In the MPC, let $H_p$ 
and $\mathcal{K}_p=\{1,..,N_p\}$ denote the prediction horizon and the prediction control steps,  where $T_p$ is the duration of each control period in the slow timescale and $N_p=H_p/{T}_p$.} 

{Define the parameters and variables in the column vector form: let $\bm{n}_{tap}(t)=[{n}_{tap}^{\phi}(t)]_{{\phi\in\Phi_{0}}}$,$\Delta\bm{n}_{tap}(t)=[\Delta{n}_{tap}^{\phi}(t)]_{{\phi\in\Phi_{0}}}\in\mathbb{Z}^{3}$ denote the tap position and change of OLTC, $\bm{n}_{cb}(t)=[\bm{n}_{i,cb}(t)]_{i\in\mathcal{N}}$,$\Delta\bm{n}_{tap}(t)=[\Delta\bm{n}_{i,cb}(t)]_{i\in\mathcal{N}}\in\mathbb{Z}^{m}$ denote the number and switching times of CBs, 
$\bm{q}^l(t)=[\bm{q}_i^l(t)]_{i\in{\mathcal{N}}}$,  $\bm{q}^{cb}(t)=[\bm{q}_i^{cb}(t)]_{i\in{\mathcal{N}}}\in\mathbb{R}^{m}$ denote the reactive load consumption power, reactive power generated by CBs, respectively. Let $\Delta\bm{q}^{cb}=[\Delta\bm{q}_i^{cb}]_{i\in\mathcal{N}}\in\mathbb{R}^{m}$ denote capacitor bank unit reactive power output, and $\Delta{tap}=[\Delta^{\phi}{tap}]_{\phi\in\Phi_{0}}\in\mathbb{R}^3$ denote the tap step size of OLTC, where $\Delta^{\phi}{tap}$ is the tap step size of OLTC in phase $\phi$. Then $v_0^{\phi}(t)$ can be  expressed as follows:
\begin{equation}\label{eq:NonlinearTap}
    {v}_{0}^{\phi}(t)=(1+n^{\phi}_{tap}(t)\Delta^{\phi}{tap})^2,\forall{t}\in\mathcal{K}_p
\end{equation}
However, (\ref{eq:NonlinearTap}) could lead to a nonlinear and nonconvex mixed-integer optimization problem, making the optimization problem hard to solve. To this end,  a linear approximation of  (\ref{eq:NonlinearTap}) is derived as follows:
\begin{equation}
\begin{split}
    {v}_{0}^{\phi}(t)&=1+2n^{\phi}_{tap}(t)\Delta^{\phi}{tap}+(n^{\phi}_{tap}(t)\Delta^{\phi}{tap})^2\\
    &\approx 1+2n^{\phi}_{tap}(t)\Delta^{\phi}{tap}, \forall{t}\in\mathcal{K}_p
\end{split}
\end{equation}
Such an approximation is believed to hold since the term $(n^{\phi}_{tap}(t)\Delta^{\phi}{tap})^2<<1$, which can be written in a compact form:
\begin{equation}\label{eq:v0}
    \bm{v}_{0}(t)=\bm{1}_{3}+2\text{diag}\big(\bm{n}_{tap}(t)\big)\Delta{tap}, \forall{t}\in\mathcal{K}_p
\end{equation}
where $\bm{1}_3$ is a $3\times{1}$ column vector of ones.}

{Finally, in the upper layer, the MPC problem can be formulated as follows:
\begin{subequations}\label{eq:UpperLayer}
\begin{align}
    \nonumber\min \sum_{t\in\mathcal{K}_p}&\frac{1}{2}\Big(||\bm{v}(t)-\bm{v}_r||_{\bm{C}_v}^2+||\bm{n}_{tap}(t)-\bm{n}_{tap}(t-1)||_{\bm{C}_{tap}}^2\\
    &+||\bm{n}_{cb}(t)-\bm{n}_{cb}(t-1)||_{\bm{C}_{cb}}^2\Big)\\
    &\text{subject to:~} (\ref{eq:CompactPFForm}) \text{~and~} (\ref{eq:v0}), \forall{t}\in\mathcal{K}_p\\
    &\bm{q}(t)=\bm{q}^c(t)-\bm{q}^g(t), \forall{t}\in\mathcal{K}_p\\
    &\bm{q}^c(t)=\bm{q}^l(t)-\bm{q}^{cb}(t), \forall{t}\in\mathcal{K}_p\\
    &\bm{\underline{q}}^g(t)\leq\bm{{q}}^g(t)\leq\bm{\overline{q}}^g(t), \forall{t}\in\mathcal{K}_p\\
    &\bm{\underline{n}}_{tap}\leq\bm{n}_{tap}(t)\leq\bm{\overline{n}}_{tap}, \forall{t}\in\mathcal{K}_p\\
    &\Delta\bm{\underline{n}}_{tap}\leq\Delta\bm{n}_{tap}(t)\leq\Delta\bm{\overline{n}}_{tap}, \forall{t}\in\mathcal{K}_p\\
    &\bm{0}\leq\bm{n}_{cb}(t)\leq\bm{\overline{n}}_{cb}, \forall{t}\in\mathcal{K}_p\\
    &\Delta\bm{\underline{n}}_{cb}\leq\Delta\bm{n}_{cb}(t)\leq\Delta\bm{\overline{n}}_{cb}, \forall{t}\in\mathcal{K}_p\\
    &\bm{q}^{cb}(t)=\text{diag}(\bm{n}_{cb}(t))\Delta{\bm{q}}^{cb}, \forall{t}\in\mathcal{K}_p
\end{align}
\end{subequations}
In (\ref{eq:UpperLayer}a), $\bm{v}_r$ is the reference of squared voltage magnitude, $\bm{C}_v$, $\bm{C}_{tap}$, and $\bm{C}_{cb}$ are the weighting matrices corresponding to the three terms. (\ref{eq:UpperLayer}b) denotes the power flow constraints and the constraints of head bus 0; (\ref{eq:UpperLayer}c)-(\ref{eq:UpperLayer}d) are the reactive power relationships; (\ref{eq:UpperLayer}e) is the VAr limits for DERs; (\ref{eq:UpperLayer}f) denotes the OLTC tap position limits; (\ref{eq:UpperLayer}g) constrains the
OLTC tap change;  (\ref{eq:UpperLayer}h) and (\ref{eq:UpperLayer}i) constrain the maximum number and switching times of CBs; (\ref{eq:UpperLayer}j) obtains the reactive power generated by CBs based on $\bm{n}_{cb}(t)$. The above problem (\ref{eq:UpperLayer}) is a  mixed-integer convex programming  (MICP) problem  and can thus be efficiently solved by MICP solvers. Only the optimal solution of OLTC and CBs in the first time step will be implemented.
}

\begin{IEEEbiography}[{\includegraphics[width=1in,height=1.25in,clip,keepaspectratio]{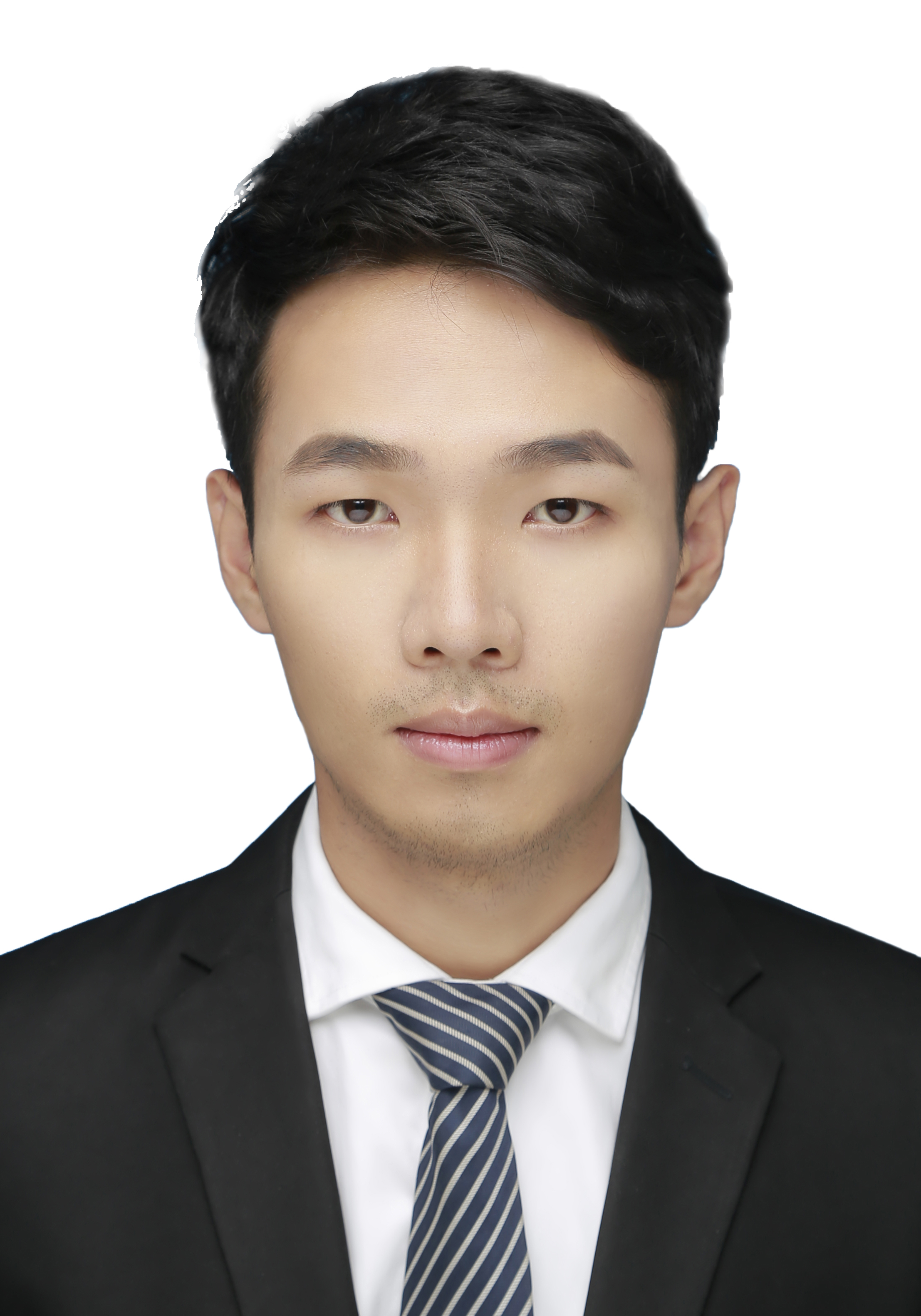}}]{Rui Cheng}(Graduate Student Member, IEEE)
is currently a Ph.D.\ student in the Department of Electrical \& Computer Engineering at Iowa State University.
He received the B.S.\ degree in electrical engineering from Hangzhou Dianzi University in 2015 and the M.S.\ degree in electrical engineering from North China Electric Power University in 2018. From 2018 to 2019, he was an Electrical Engineer with State Grid Corporation of China, Hangzhou, China.

His research interests include power distribution systems, voltage/var control, transactive energy markets, and applications of optimization and machine learning methods to power systems.
\end{IEEEbiography}
\begin{IEEEbiography}[{\includegraphics[width=1in,height=1.25in,clip,keepaspectratio]{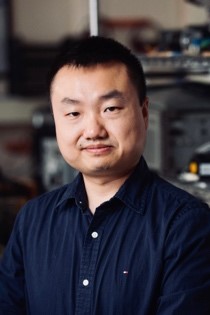}}]{Zhaoyu Wang} (Senior Member, IEEE) received the B.S. and M.S. degrees in electrical engineering from Shanghai Jiaotong University, and the M.S. and Ph.D. degrees in electrical and computer engineering from Georgia Institute of Technology. He is the Northrop Grumman Endowed Associate Professor with Iowa State University. His research interests include optimization and data analytics in power distribution systems and microgrids. He was the recipient of the National Science Foundation CAREER Award, the Society-Level Outstanding Young Engineer Award from IEEE Power and Energy Society (PES), the Northrop Grumman Endowment, College of Engineering’s Early Achievement in Research Award, and the Harpole-Pentair Young Faculty Award Endowment. He is the Principal Investigator for a multitude of projects funded by the National Science Foundation, the Department of Energy, National Laboratories, PSERC, and Iowa Economic Development Authority. He is the Chair of IEEE PES PSOPE Award Subcommittee, the Co-Vice Chair of PES Distribution System Operation and Planning Subcommittee, and the Vice Chair of PES Task Force on Advances in Natural Disaster Mitigation Methods. He is an Associate Editor of IEEE TRANSACTIONS ON POWER SYSTEMS, IEEE TRANSACTIONS ON SMART GRID, IEEE OPEN ACCESS JOURNAL OF POWER AND ENERGY, IEEE POWER ENGINEERING LETTERS, and IET Smart Grid.
\end{IEEEbiography}
\begin{IEEEbiography}[{\includegraphics[width=1in,height=1.25in,clip,keepaspectratio]{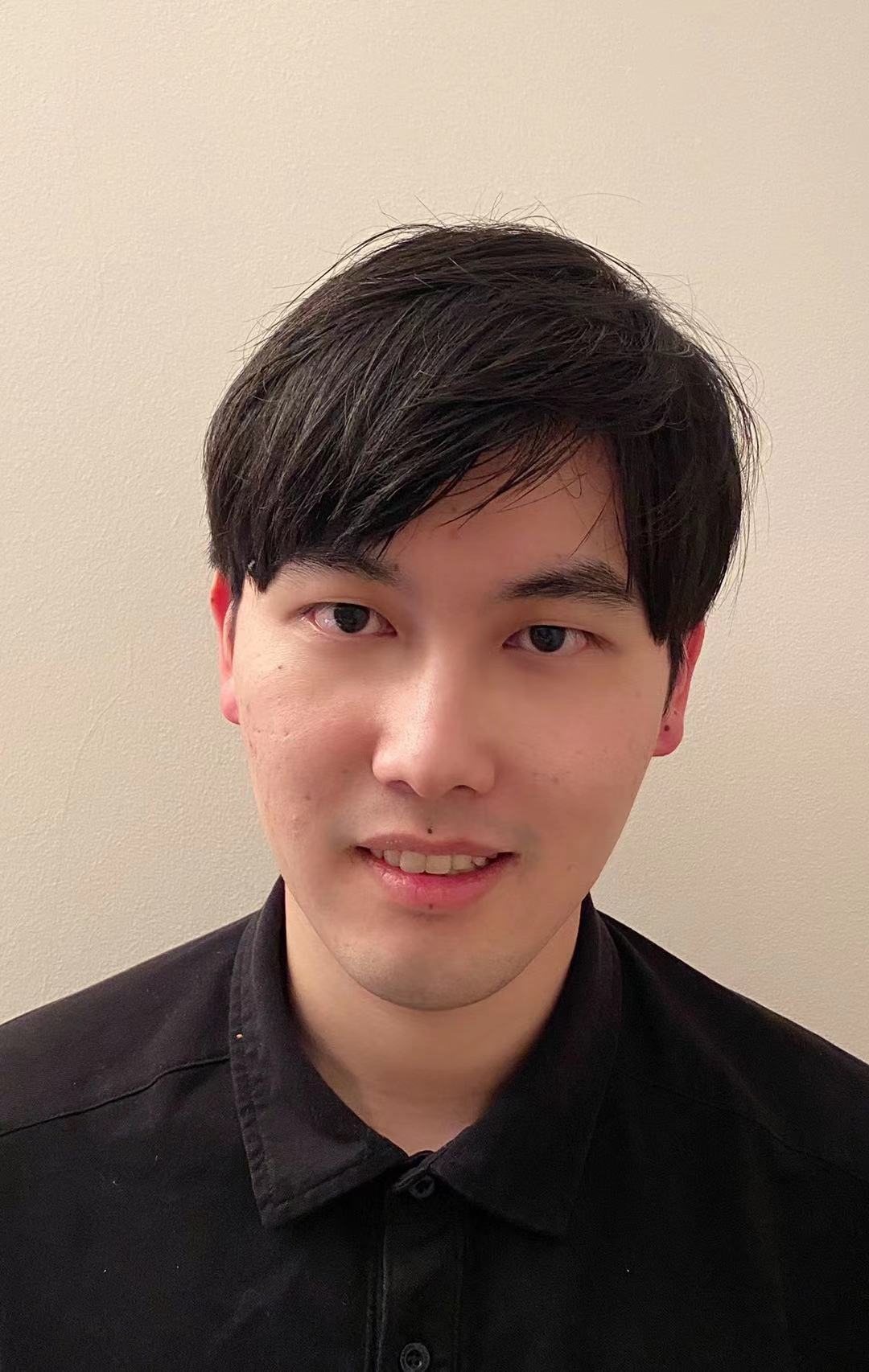}}]{Yifei Guo}(M'19) received the B.E. and Ph. D. degrees in electrical engineering from Shandong University, Jinan, China, in 2014 and 2019, respectively.
Currently, he is a Postdoctoral Research Associate with the Department of Electrical and Computer Engineering,  Iowa State University, Ames, IA, USA. He was a visiting student with the Department of Electrical Engineering, Technical University of Denmark, Lyngby, Denmark, in 2017--2018. 

His research interests include voltage/var control, renewable energy integration, wind farm control, distribution system optimization and control, and power system protection.
\end{IEEEbiography}
\begin{IEEEbiography}[{\includegraphics[width=1in,height=1.25in,clip,keepaspectratio]{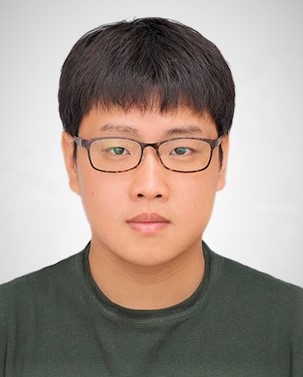}}]{Qianzhi Zhang} (Graduate Student Member, IEEE) received the M.S. degree in electrical engineering from Arizona State University in 2015. He is currently pursuing the Ph.D. degree with the Department of Electrical and Computer Engineering, Iowa State University, Ames, IA, USA. He was a research engineer with Huadian Electric Power Research Institute from 2015 to 2016. He was a recipient of the Outstanding Reviewer Award from IEEE Transaction on Smart Grid, IEEE Transaction on Power Systems, and International Journal of Electrical Power and Energy Systems. His research interests include voltage/var control, power/energy management, system resilience enhancement, application of advanced optimization and machine learning techniques in power system operation and control.  
\end{IEEEbiography}
\end{document}